\documentclass[aps,prb,twocolumn, superscriptaddress]{revtex4-2}

\usepackage{graphicx}

\usepackage{multirow}
\usepackage{float}

\usepackage{color} 

\begin{document}

\preprint{AAPM/123-QED}

\title{Signatures of Orbital Order and Disorder in Fluoro-Perovskites with \textit{t$_{2g}$} Electronic Degeneracies}

\author{C. A. Crawford} 
\affiliation{Department of Chemistry, University of Warwick, Gibbet Hill, Coventry, CV4 7AL, UK}

\author{C. I. Hiley} 
\affiliation{Department of Chemistry, University of Warwick, Gibbet Hill, Coventry, CV4 7AL, UK}

\author{J. Gainza}
\affiliation{European Synchrotron Radiation Facility, 71 Avenue des Martyrs, CS40220, 38043 Grenoble Cédex 9, France}

\author{C. Ritter}
\affiliation{Institut Laue-Langevin, 71 Avenue des Martyrs, CS20156, 38042 Grenoble Cédex 9, France}

\author{R. I. Walton} 
\email{r.i.walton@warwick.ac.uk}
\affiliation{Department of Chemistry, University of Warwick, Gibbet Hill, Coventry, CV4 7AL, UK}

\author{M. S. Senn}
\email{m.senn@warwick.ac.uk}
\affiliation{Department of Chemistry, University of Warwick, Gibbet Hill, Coventry, CV4 7AL, UK}

\date{\today}

\begin{abstract}
A detailed high-resolution, variable temperature powder diffraction study of the fluoro-perovskites NaFeF$_3$ and NaCoF$_3$ is performed to probe their orbital ordering transitions. Through analysis of the symmetry adapted macrostrains and atomic distortions, we show that NaFeF$_3$ undergoes a C-type orbital order transition associated with the \textit{t$_{2g}^4$} states of Fe$^{2+}$. Counter-intuitively, the phase transition leading to the orbital order appears second order-like, which contradicts the thermodynamic requirements for electronic and isosymmeric phase transitions, implying that there must be an associated hidden symmetry breaking. On the other hand, for NaCoF$_3$, consideration of the symmetry adapted strains allows us to confidently rule out the occurrence of any long-range orbital orders down to 4 K. Since NaCoF$_3$ is an insulator with quenched orbital angular momentum at this temperature, our findings point towards a novel kind of orbital disorder associated to the \textit{t$_{2g}^5$} electronic degeneracy. 
\end{abstract}

\maketitle

\section{\label{sec:level1}Introduction}

\begin{figure*}
\includegraphics[scale=1]{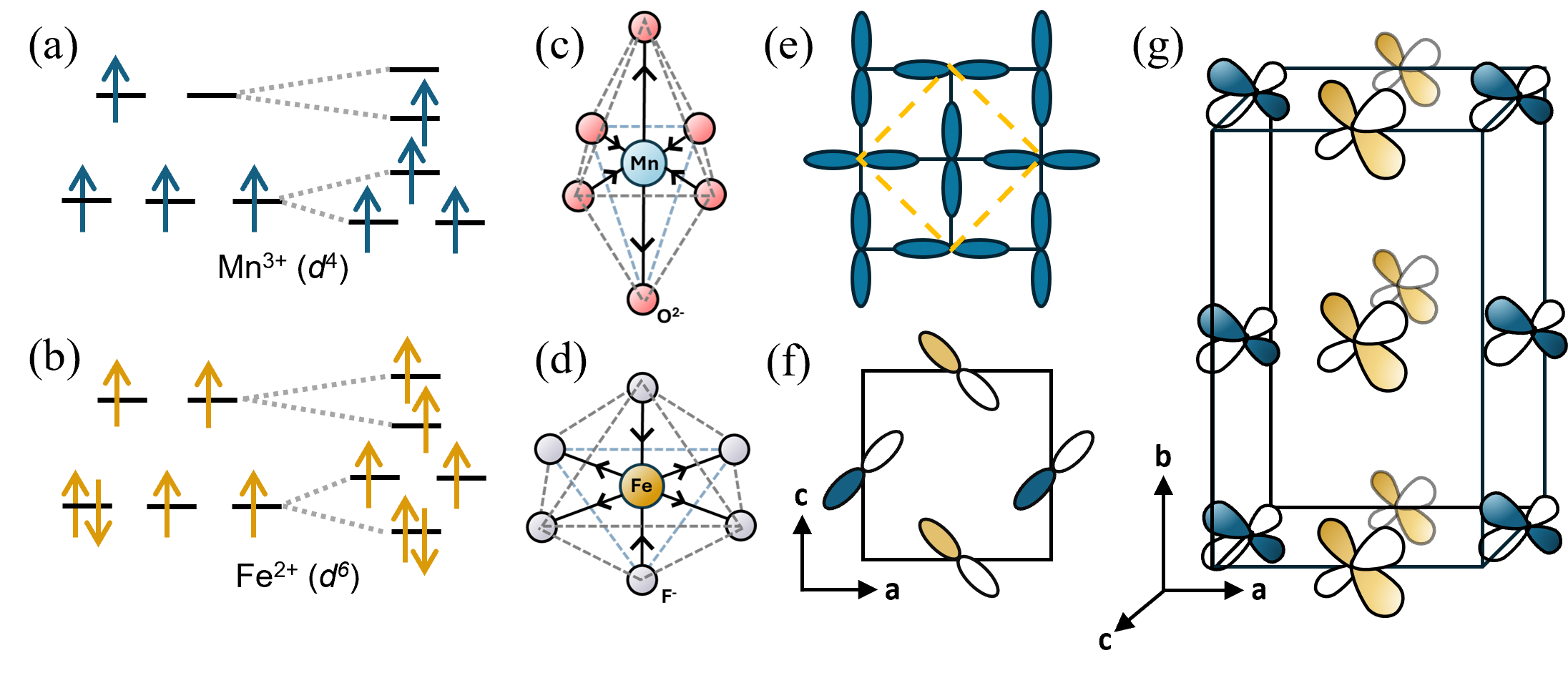}
\caption{\label{fig1}Crystal field splitting diagrams showing the removal of degeneracy of the \textbf{a}. \textit{e$_g$} and \textbf{b}. \textit{t$_{2g}$} orbitals. These distortions manifest as an \textbf{c}. elongation (MnO$_6$) and \textbf{d}. compression (FeF$_6$) of the \textit{M}F$_6$ octahedra respectively.  \textbf{e}. Schematic showing orbital order of the \textit{d$_{z2}$} orbitals in a 2 $\times$ 2 primitive cubic cell. Dashed lines denote the new unit cell to describe the OO. \textbf{f}, \textbf{g}. C-type orbital order of the \textit{d$_{xy}$} orbitals in \textit{Pnma}.}
\end{figure*}

For decades oxide-based perovskites, ABO$_3$, have been subject to thorough investigation regarding their extensive structural phase transitions and subsequent properties \cite{Mitchell2002Perovskites:Ancient}. The occurrence of symmetry-lowering phase transitions with decreasing temperature can be driven by a multitude of mechanisms, including the onset of various ordering schemes such as magnetic order or charge ordering, or by the variation and/or onset of octahedral rotations related to the tolerance factor (i.e. the relationship between the size of the A and B site cations and the anion). Structural transitions may also be electronically driven depending on the cations present and result in changes to the coordination environments of these cations. A well-studied phenomenon is the Jahn-Teller (JT) distortion that results from the lifting of degeneracy of the \textit{e$_g$} orbitals that have an unpaired spin (Fig.\ \ref{fig1}(a)). This drives the divergence of bond lengths within an MX$_6$ octahedron which results in either a compression or elongation (Fig.\ \ref{fig1}(c,d)). Over a periodic structure, a long-range orbital ordering (OO) of cooperative JT distortions can manifest to reduce the coulombic repulsion between neighbouring atoms and reduce the total energy of the system \cite{Lufaso2004Jahn-TellerPerovskites}\cite{Streltsov2017OrbitalTrends}. Where degeneracy has been lifted from the \textit{e$_g$} orbital (e.g. Cu$^{2+}$ or Mn$^{3+}$) an ordering of the \textit{d$_{z^2}$} orbital along various \textit{x}, \textit{y} and \textit{z} axes can occur (Fig.\ \ref{fig1}(e)). Orbital order arising from cooperative JT distortions has been observed in both oxide (LaMnO$_3$) \cite{Goodenough2007OrbitalPerovskites} and fluoride (NaCuF$_3$ \cite{Bernal2020StructuralNaCrF3} and KCuF$_3$ \cite{Zhou2011Jahn-TellerPressure}) perovskites. When considering an aristotype, cubic perovskite cell, ordering of the \textit{e$_g$} \textit{d$_{z^2}$}  orbitals results in a symmetry-lowering expansion of the unit cell (denoted by the dashed line in Fig.\ \ref{fig1}(e)) leading to the observation of supercell reflections in diffraction data. In C-type OO, the orbitals order within “layers” along a specified axis, for example in the case of LaMnO$_3$, when crystallising in space group \textit{Pnma}, the orbitals are ordered in the \textit{ac} plane, with no alternating orbital order along \textit{b}. This can be observed crystallographically by the alternating directions of the elongation of the MnO$_6$ octahedra evidenced by variations in the bond lengths. In LaMnO$_3$ and A-site substituted LaMnO$_3$  perovskites, at the onset of orbital order, the resistivity of the material drops in a metal-to-insulator transition \cite{Tragheim2024ProximityPerovskites}\cite{Pissas2005Crystal0.11x0.175}, meaning the study of compositionally simple perovskites may provide further insight and implications for the tuning of functional properties such as colossal magnetoresistance which is directly related to orbital order and disorder.  
 
JT distortions arising from electronic degeneracies in the \textit{t$_{2g}$} orbitals in perovskites are also possible, though have not been studied to the same extent as JT distortions of the \textit{e$_g$} orbitals \cite{Zhang2020OriginLaTiO3}. Typically, distortions arising from the \textit{e$_g$} orbitals result in bond length variations whereas local distortion modes exist that can remove the degeneracy of the \textit{t$_{2g}$} orbitals (Fig.\ \ref{fig1}(b)) via bond angle shearing of the octahedra. In practice, similar bond length distortions, although of much smaller magnitudes compared to those arising from \textit{e$_g$} degeneracy, are also found to be associated with the removal of \textit{t$_{2g}$} orbital degeneracy and long range OO in these systems. In the LnVO$_3$ series (Ln = lanthanoide), V$^{3+}$ (\textit{d}$^2$) undergoes various orbital ordering transitions depending on the size of the lanthanoide ion. For all Ln cations, the VO$_6$ arrangement undergoes a G-type OO transition from \textit{Pnma} to \textit{P}2$_1$/\textit{b}11, and those where Ln = Dy to Lu and Y, have a further transition back to \textit{Pnma} which has a C-type orbital order \cite{Varignon2015CouplingPerovskites}\cite{Johnson2012X-raySmVO3}\cite{Sage2007CompetingExpansion}. This series of phase transitions has also been observed in TmVO$_3$ \cite{Ritter2016CrystallographicTmVO3}. The magnitude of JT distortions of the \textit{t$_{2g}$} orbitals are far smaller than those of the \textit{e$_g$} orbitals, and so care must be taken not to mistake the additional degrees of freedom afforded by the symmetry lowering in orthorhombic perovskites (i.e. octahedral rotations) for those arising from electronic effects. With this in mind one might look to the other transition-metal (TM) cations that may have \textit{t$_{2g}$} JT activity, such as Co$^{3+}$ and Fe$^{2+}$ (\textit{d}$^6$), Ni$^{3+}$ (\textit{d}$^7$). However, the LnCoO$_3$ (\textit{d}$^6$) generally remains in the low spin state with all \textit{t$_{2g}$} orbitals doubly occupied and no long-range Co magnetic order is observed \cite{Bull2016Low-temperatureStudy}; and LnNiO$_3$ (\textit{d}$^7$) may charge disproportionate to Ni$^{2+}$ and Ni$^{4+}$, where the charge ordering directly competes with and interrupts any long-range OO. Additionally, many AFe$^{2+}$X$_3$ materials do not have the perovskite structure but rather form isolated 1D edge-sharing chains of octahedra (KFeCl$_3$, KFeBr$_3$ \cite{Amit1974PreparationKFeBr3}) or face-sharing chains (CsFeCl$_3$ \cite{Kohne1993CrystalCsFeC3}, CsFeBr$_3$ \cite{Takeda1974StructureCsFeBr3}) with no evidence of JT distortions\textbf. 

Switching the anion from oxide to fluoride, provides an opportunity to study the late \textit{d}-block transitions metals in their 2+ oxidation state with electron configurations that would result in electronic degeneracies of the \textit{t$_{2g}$} rather than \textit{e$_{g}$} orbitals. With this in mind, we have synthesised NaCoF$_3$ (\textit{d}$^7$) and NaFeF$_3$ (\textit{d}$^6$). At room temperature, both compositions crystallise in the GdFeO$_3$ archetype, \textit{Pnma} space group. A recent study on NaFeF$_3$ determined the magnetic structure but did not resolve the existence of orbital order \cite{Bernal2020CantedMethod}. Here we seek to compare the behaviour between NaFeF$_3$ and NaCoF$_3$, where a naïve interpretation of the \textit{t$_{2g}$} occupancies would lead to an expectation of JT distortions characterised by different patterns of bond length compressions (Fe, Fig.\ \ref{fig1}(b)) and elongations (Co, equivalent splitting to Mn$^{3+}$ Fig.\ \ref{fig1}(a)) in the two compounds. Our study aims to deconvolute the bond length distortions that arise from JT distortions and those that are intrinsic to the \textit{Pnma} perovskite system by comparing the magnitude of structural distortions with two further examples, Na\textit{M}F$_3$ (\textit{M} = Mn, Ni) which are expected to be JT-inactive.

\section{\label{sec:level1}Methods}
NaF (Acros Organics, 97\%), MnCl$_2$·4H$_2$O (VWR, analytical grade), FeCl$_2$·4H$_2$O, (Thermo Scientific, 99+ \%), CoCl$_2$·6H$_2$O (Thermo Scientific, 99.9+ \% ), NiCl$_2$·6H$_2$O (Sigma Aldrich, 99\%) and ethylene glycol (Fisher, $\geq$ 99\%) were all used as received without further purification. Polycrystalline powder samples were synthesised using a solvothermal method, whereby 0.060 mol NaF (2.519 g) and 0.020 mol \textit{M}Cl$_2$.xH$_2$O (\textit{M} = Mn, Fe, Co, Ni) were heated in 100 ml of ethylene glycol in a sealed 200 ml autoclave at 120 °C for 120 hours. The samples were filtered and dried at 70 °C overnight to give approximately 2.5 g of perovskite. If initial powder diffraction showed unreacted NaF was present, the samples were stirred in 10 mL distilled water for 24 hours before being re-filtered and again dried at 70 °C overnight (this was not done for NaFeF$_3$, to prevent oxidation to Na$_3$FeF$_6$). The powders were initially characterised by powder X-ray diffraction (PXRD) on a Panalytical Empyrean diffractometer with a Cu K$\alpha$$_{1,2}$ ($\lambda$ = 1.5406 Å, 1.5444 Å) source to confirm phase purity. 

Magnetometry measurements were performed on a Quantum Design MPMS-5S SQUID magnetometer. Around 20 mg of accurately weighed powdered sample was loaded into a gel capsule, encased within a plastic straw. Zero field cooled (ZFC) and field cooled (FC) DC magnetic susceptibility  measurements were taken in the range of 2 – 300 K with an applied field of 1000 Oe . 

High resolution synchrotron PXRD measurements were performed at beamline ID22 ($\lambda$ = 0.35433788(8) Å) at the European Synchrotron Radiation Facility (ESRF) between 4 K – 300 K using a Dynaflow ESRF cryostat \cite{Fitch2023ID22ESRF}. Further variable temperature PXRD data were collected between 100 K to 500 K at beamline I11 ($\lambda$ = 0.825257 (7) Å) at Diamond Light Source. Powder neutron diffraction (PND) data for NaCoF$_3$ were collected on the high-resolution diffractometer, D2B ($\lambda$ = 1.595 Å) at the Institut Laue Langevin (ILL) with samples loaded into a 9 mm cylindrical vanadium can. 

Data were analysed by the Rietveld method using TOPAS Academic, version 7 \cite{Coelho2017TOPAS-Academic} with an incorporated jEdit text editor. Using files generated by ISODISTORT \cite{Campbell2006ISODISPLACE:Distortions}\cite{StokesISODISTORTSuite}\cite{StokesINVARIANTSIso.byu.edu.}\cite{Hatch2003INVARIANTS:Group}, a symmetry-based approach was used to investigate structural distortions. Crystallographic information files (CIF) were visualised using structure viewing software, VESTA \cite{Momma2011VESTA3Data}. The Python package, 
 \textit{VanVleckCalculator} \cite{Nagle-Cocco2024VanVanVleckCalculator} was used to aid the analysis of Van Vleck distortion modes quantifying the distortion of the B-site octahedra. Details on set up of the octahedral axes are provided in S2 \cite{Crawford2025SignaturesMaterial}.   

\section{\label{sec:level1}Results and Discussion}

Initial PXRD measurements at room temperature confirmed phase formation for all Na\textit{M}F$_3$ perovskites with only minor NaF impurities present. Rietveld models at 10 K using data collected at ID22 are provided in Fig.\ \ref{fig:S1} \cite{Crawford2025SignaturesMaterial}. Magnetometry results (provided in S6 \cite{Crawford2025SignaturesMaterial}) for all materials show bulk antiferromagnetic (AFM) transitions accompanied by a weak ferromagnetic (FM) component as evidenced by the splitting of the field cooled (FC) and zero-field cooled (ZFC) measurements. Curie-Weiss fitting of the paramagnetic region for \textit{M} = Fe and Co yielded values that were indicative of both transition metals being in their high-spin electron configuration, in line with previous reports \cite{Bernal2020CantedMethod}. This is important to establish when comparing NaFeF$_3$ to LnCoO$_3$, since while both are \textit{d}$^6$, Co$^{3+}$ tends to form the low spin \cite{Bull2016Low-temperatureStudy}, \textit{t$_{2g}^6$} electron configuration which is neither JT active nor magnetic.

\begin{figure}
\includegraphics[scale=1]{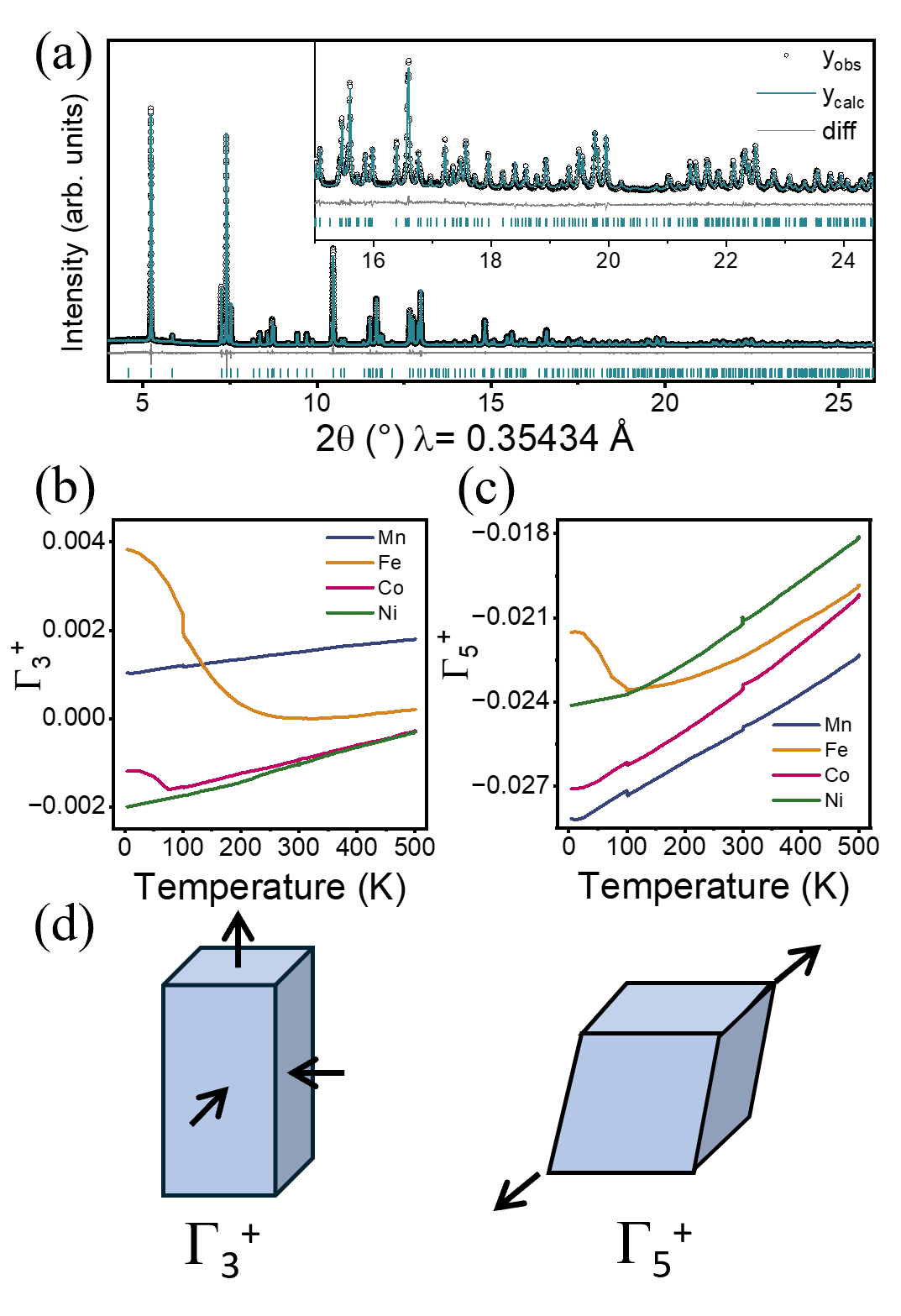}
\caption{\label{fig2}\textbf{a}. Rietveld refinement of NaFeF$_3$ in \textit{Pnma} using XRD data collected at ID22 ($\lambda$ = 0.35433788 (8) Å) at 4 K. \textbf{b}. $\Gamma_3^+$ and \textbf{c}. $\Gamma_5^+$ strain modes determined from symmetry mode refinements for Na\textit{M}F$_3$. \textbf{d}. schematic depictions of the $\Gamma_3^+$ and $\Gamma_5^+$ strain modes.}
\end{figure}

Rietveld refinements carried out against the synchrotron X-ray data (both ID22 and I11) show that all Na\textit{M}F$_3$ crystallise in \textit{Pnma} between 4 K to 500 K. By performing distortion mode refinements generated using ISODISTORT, it is possible to extract further information on the symmetry lowering structural distortions away from the aristotype \textit{Pm}$\bar{3}$\textit{m}  perovskite via irreducible representation (irrep) analysis. Within this formalism the symmetry breaking strain modes, related to the refined lattice parameters transform as irreps $\Gamma_3^+$ and $\Gamma_5^+$ (Fig.\ \ref{fig2}(b,c,d)). Symmetrised strains for \textit{M} = Co and Fe show a deviation from the expected trend at lower temperatures hinting at possible phase transitions, when compared to \textit{M} = Ni and Mn where no OO is expected. While the smaller deviation in $\Gamma_3^+$ in Co is concomitant with its AFM transition temperature, \textit{T$_N$} (78(1) K) and therefore likely related to magnetostriction, the deviation in $\Gamma_3^+$ for Fe (\textit{T}$_N$ = 89(1) K) occurs far above this, at nearer 300 K  (Fig.\ \ref{fig2}(b)), and is significantly larger in magnitude suggesting it is not of magnetostricitve origin. In the basis of the \textit{Pnma} crystal structure, this has a pronounced effect in the \textit{b} lattice parameter (Fig.\ \ref{fig:S1} \cite{Crawford2025SignaturesMaterial}) which shows a clear negative thermal expansion below 160 K, and plateaus below \textit{T$_N$}. Bernal \textit{et al}. observed similar behaviour in the lattice parameter trends for NaFeF$_3$, however they ascribed this to magnetostriction based on the observation of a diffuse magnetic peak occurring below 120 K \cite{Bernal2020CantedMethod}. This is well below the onset of the rapid increase of $\Gamma_3^+$ strain shown in Fig.\ \ref{fig2}(b) but may be related to the small anomaly we see in $\Gamma_5^+$ at this temperature (Fig.\ \ref{fig2}(c)), highlighting the benefit of working in this symmetry-adapted basis.

\begin{figure}
\includegraphics[scale=0.9]{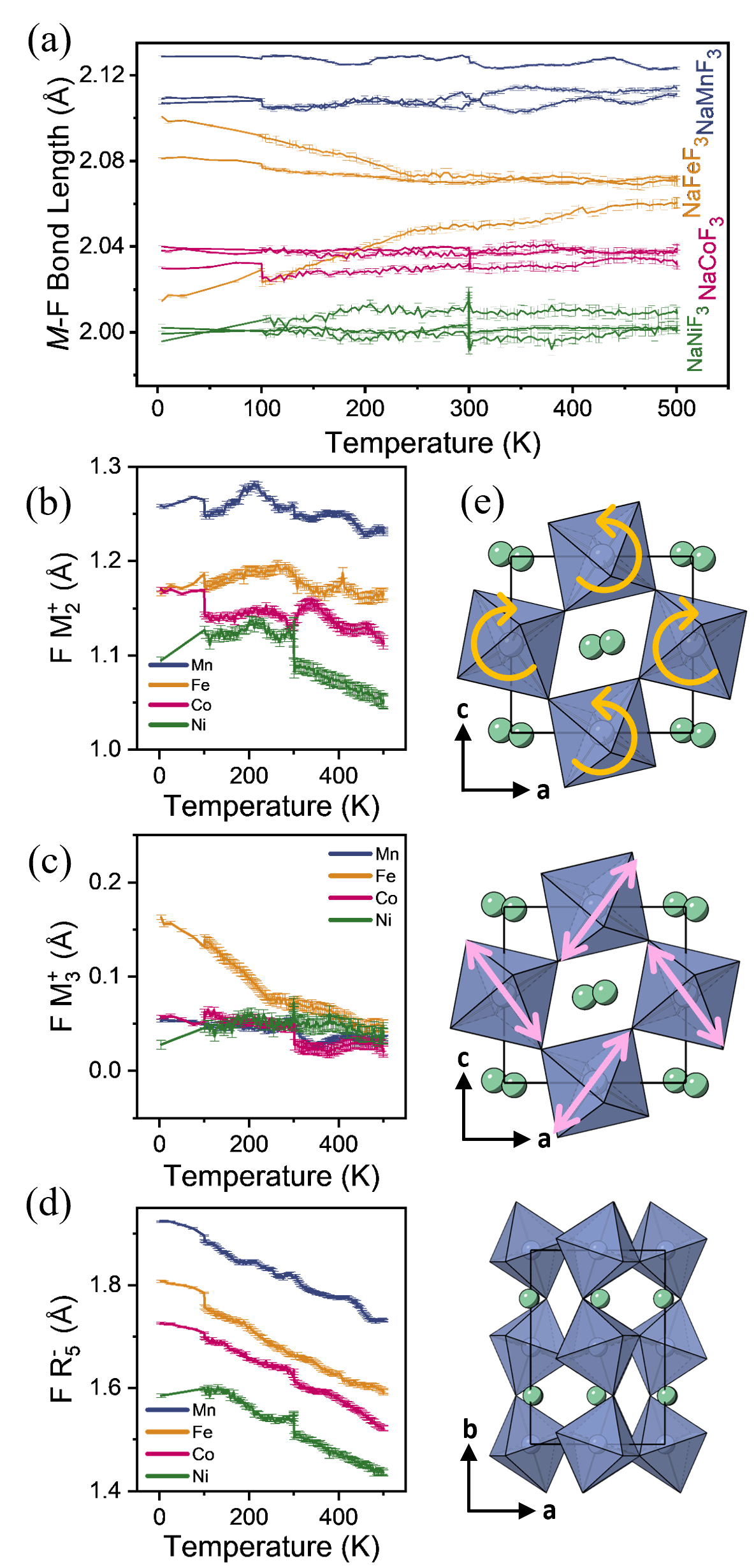}
\caption{\label{fig3}\textbf{a}. B-F bond lengths versus temperature highlighting the octahedral compression in NaFeF$_3$ compared to Mn, Co and Ni. Variations in the; \textbf{b}. M$_2^+$, \textbf{c}. M$_3^+$ and \textbf{d}. R$_5^-$ distortion modes with temperature, determined from symmetry mode refinements. \textbf{e}. depictions of the (top to bottom) M$_2^+$, M$_3^+$ and R$_5^-$ distortion modes.}
\end{figure}

Fig.\ \ref{fig3}(a) shows the \textit{M}-F bond lengths with decreasing temperature. While it is noted that in \textit{Pnma} there is an allowed deviation of the bond lengths from each other, the deviation observed in NaFeF$_3$ is far greater than any other in the series and becomes more pronounced with decreasing temperature. It also shows the  “four long, two short” arrangement as expected when considering the crystal-field splitting in Fig.\ \ref{fig1}(b).

As well as providing an orthogonal basis in which to view the evolution of the structural degrees of freedom, when using symmetry mode analysis, in many cases the irreps have a fortuitous direct correspondence with physically intuitive origins such as octahedral tilting, bond length distortions and antipolar displacements \cite{Senn2018APerovskites}\footnote{All irrep labels are given with respect to a \textit{Pm}$\bar{3}$\textit{m} unit cell with the A site at the origin.}. The M$_2^+$ and R$_5^-$ distortion modes describe the in-phase and out-of-phase tilting of the B-site octahedra respectively, whereas  M$_3^+$ describes B-X bond length distortions associated with a JT elongation/compression of the octahedra  which are shown in Fig.\ \ref{fig3}(c). The trends (Fig.\ \ref{fig2}(b,d)) in NaFeF$_3$ for the M$_2^+$ and R$_5^-$  mode amplitudes follow those of the other transition metals, however the M$_3^+$ mode which controls the octahedral compression/elongation drastically increases with decreasing temperature to be over three times as large at 4 K than for the other transition metals. All additional distortion modes are provided in Fig.\ \ref{fig:S2}  \cite{Crawford2025SignaturesMaterial}.

To further understand the underpinning distortions, the Van Vleck distortion modes (\textit{Q} modes) \cite{VanVleck1939TheXY6} for the MX$_6$ octahedra were calculated. \textit{Q}$_2$ and \textit{Q}$_3$ describe bond length distortions, and were calculated using bond lengths whereas \textit{Q}$_{4-6}$ describe bond angle distortions (Fig.\ \ref{fig:S5} \cite{Crawford2025SignaturesMaterial}). To calculate \textit{Q}$_{4-6}$, the Python package, \textit{VanVleckCalculator} \cite{Nagle-Cocco2024VanVanVleckCalculator} was used. This method allows the \textit{Q} modes to be extracted without disregarding any angular distortion resulting from shearing of the octahedra; a point we consider important to investigate since in principle, these shearings provide an alternative mechanism for removing the degeneracy of the local \textit{t$_{2g}$} electronic state. \textit{Q}$_1$ is a breathing mode, and \textit{Q}$_{2,3}$ have transformational \textit{E$_g$} symmetry, whereas \textit{Q}$_{4-6}$ distortions are of \textit{T$_{2g}$} symmetry \cite{VanVleck1939TheXY6}. Thus, these \textit{Q} modes form a natural basis by which to investigate the local OO. The \textit{Q}$_2$ mode character reflects a “2 long, 2 medium, 2 short” planar rhombic distortion, and \textit{Q}$_3$ as a Jahn-Teller 2:4 tetragonal elongation or compression. \textit{Q}$_{4-6}$ do not appear to have significant variations with temperature (see Fig.\ \ref{fig:S4}) however clearly \textit{Q}$_2$ and \textit{Q}$_3$ appear to behave like an order parameter in NaFeF$_3$. This is in stark contrast to the behaviour of M$^{3+}$ mode amplitudes and \textit{Q}$_2$/\textit{Q}$_3$ distortions in \textit{M} = Mn, Co or Ni which have small, temperature independent values.

\begin{figure*}
\includegraphics[scale=0.8]{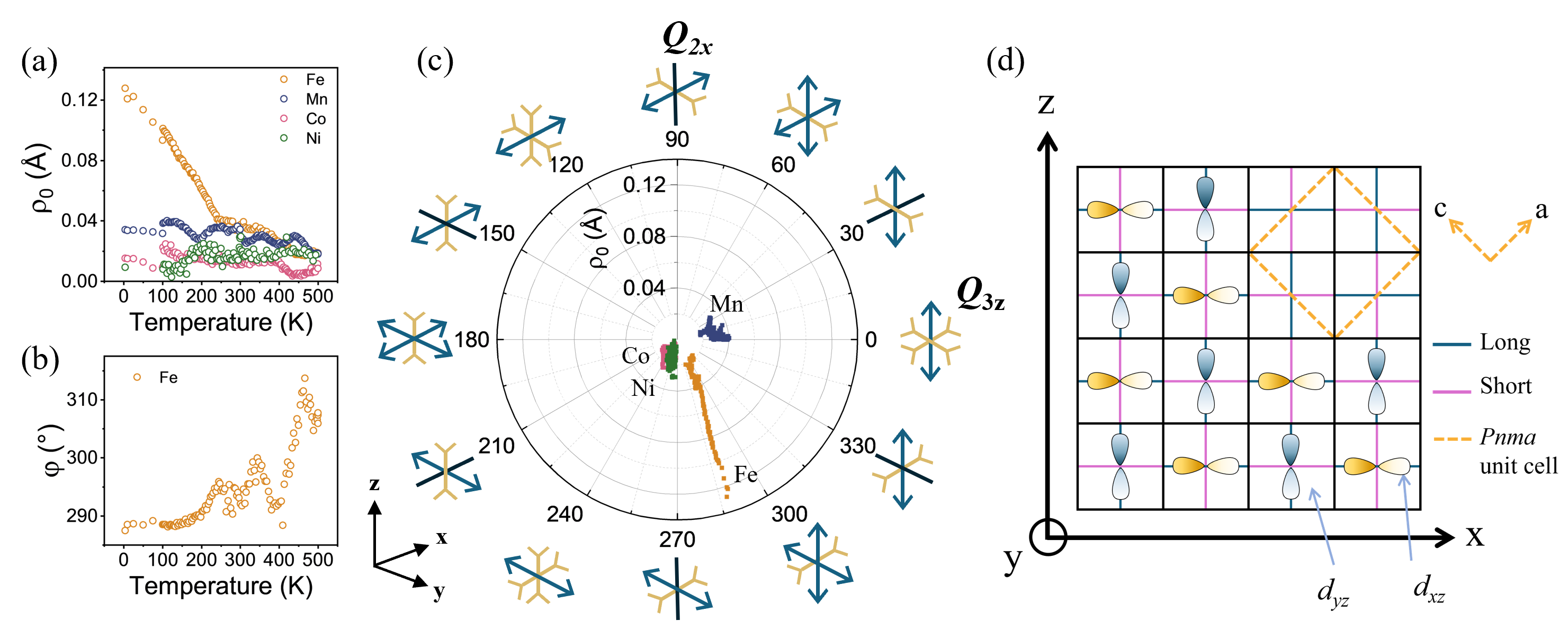}
\caption{\label{fig4}.\textbf{a}. $\rho_0$ vs temperature for Na\textit{M}F$_3$. \textbf{b}. $\phi$ vs Temperature for NaFeF$_3$.\textbf{c}. Polar plot adapted from \cite{Goodwin2017OrbitalTemperatures} showing the \textit{Q}$_2$-\textit{Q}$_3$ distortion space with the magnitudes of Na\textit{M}F$_3$ shown. The axes in the bottom left corner depict the axes of the octahedron. \textbf{d}. A schematic highlighting the ordering of the doubly filled \textit{d$_{xz}$} and \textit{d$_{yz}$} orbitals.}
\end{figure*}

Since the symmetry breaking distortions that lead to these local \textit{Q}$_{2,3}$ distortions are formally allowed through improper couplings with octahedral rotations that give rise to the observed \textit{Pnma} symmetry in these perovskites, it is important to untangle those contributions from that intrinsically driven by the electronic ordering. Here, a comparison of different compositions with similar tolerance factors but varying electronic degeneracies enable these two factors to be unpicked. To do this, the Jahn-Teller parameters $\phi$ and $\rho_0$ were calculated; $\phi$ = arctan(\textit{Q$_2$}/\textit{Q$_3$}) and $\rho_0$ = (\textit{Q$_2^2$}+\textit{Q$_3^2$})$^{1/2}$. $\phi$ and $\rho_0$ (Fig.\ \ref{fig4}(a,b)) describe the angle and vector magnitude of JT distortion respectively within the \textit{Q}$_2$-\textit{Q}$_3$ distortion space, used to describe octahedral distortions occurring due to the lifting of degeneracy within the \textit{e$_g$} orbitals. Whilst Fe$^{2+}$ has degeneracy within the \textit{t$_{2g}$} orbitals, the resulting octahedral deformations also cause a splitting of the \textit{e$_g$} orbitals, meaning that there may be signatures of these distortions within the \textit{e$_g$} phase space. A polar plot of the \textit{Q}$_2$-\textit{Q}$_3$ phase space is shown in Fig.\ \ref{fig4}(c) which describes the octahedral configuration along different octahedral directions \cite{Kanamori1960CrystalCompounds}\cite{Goodenough1998JAHN-TELLERSOLIDS}. The magnitude of the distortion observed in Fe$^{2+}$ is far greater than the other transition metals, clearly indicating the presence of the JT distortion (see Fig.\ \ref{fig4}(a)). As anticipated the tetragonal compression occurs along \textit{x}, with an elongation along \textit{y} and \textit{z}, indicative of the fully occupied, stabilised \textit{d$_{yz}$} orbital which would be consistent with the expected splitting for a \textit{d}$^6$ high spin cation. A 90$^{\circ}$ rotation of the octahedra on the neighbouring B site corresponds to a swapping of the long and short bond lengths along \textit{x} and \textit{z}, resulting in a $\phi$ angle where the compression occurs along \textit{y} meaning a stabilisation of the \textit{d$_{xz}$} orbital (Fig.\ \ref{fig:s6}\cite{Crawford2025SignaturesMaterial}). This relationship is imposed by the \textit{Pnma} symmetry and is a form of C-type orbital order. It should be noted that the $\phi$ values for NaFeF$_3$ lie somewhere between that of a pure \textit{Q}$_3$ or \textit{Q}$_2$. This is a similar intermediate situation as observed in LaMnO$_3$ which has a $\phi$ of 107$^{\circ}$ i.e. somewhere between 90$^{\circ}$ (pure \textit{Q}$_2$) and 120$^{\circ}$ (pure \textit{Q}$_3$) \cite{Kanamori1960CrystalCompounds}\cite{Goodenough2007OrbitalPerovskites}. Additionally, there is no abrupt change in the magnitude of $\rho_0$ (Fig.\ \ref{fig4}(a)) which gradually increases with decreasing temperature, indicative of a second-order phase transition. This creates an apparent contradiction since an isosymmetric phase transition (i.e. a phase transition with no space group symmetry lowering) should be first order \cite{Christy1995IsosymmetricExamples}. This is true for all processes involving an instantaneous transfer of electrons between states, for example, the phase transition from orbital ordered to disordered state in LaMnO$_3$ at around 750 K is accompanied by a sharp discontinuity in the volume \cite{Rodriguez-Carvajal1998Neutron-diffractionLaMnO3}\cite{Thygesen2017LocalLaMnO3}. This contradiction would be resolved if there was some hidden symmetry breaking associated with this transition that we are not sensitive to see with the diffraction methods used to probe the phase transition. The microstrain of our sample \textit{e}$_0$ $\approx$ 0.06 \%, as determined from our high resolution powder diffraction data collected on ID22, sets an upper limit on any such coupling to macrostrain such further symmetry breaking could impose. This is approximately an order of magnitude lower than the excess strain associated with the coupling of $\Gamma_3^+$ to the OO.

As mentioned above, on cooling, NaFeF$_3$ shows no signs of long-range symmetry lowering by diffraction and is fitted well to \textit{Pnma} at all temperatures. This does not exclude the presence of Jahn-Teller distortions with a C-type orbital ordering(Fig.\ \ref{fig4}(d)), as allowed by symmetry in \textit{Pnma} perovskite. However, identifying the onset of OO becomes more complex since \textit{Pnma} has an intrinsic distortion of the \textit{M}F$_6$ bond lengths that transforms as the M$_3^+$ irreducible representation and forms a secondary order parameter to the in-phase (M$_2^+$) and out-of-phase (R$_5^-$) octahedral tiltings, and results in a 2 long, 2 medium, 2 short \textit{M}F bond lengths, as evident in \textit{M} = Ni and Mn (Fig.\ \ref{fig3}(a)). By comparing \textit{M} = Fe to \textit{M} = Ni and Mn we can disentangle the bond length distortions arising simply due to the coupling with the primary order parameter (octahedral tilting and rotations, see Fig. \ \ref{fig:s7}), and those distortions which are driven by the orbital order itself (see S4 and S5 \cite{Crawford2025SignaturesMaterial}) Taking the excess $\Gamma_3^+$ strain due to OO as proxy for the order parameter \cite{Herlihy2025InterplayRuddlesdenPoppers}, this allows us to estimate a second order phase transition temperature \textit{T}$_{OO}$ = 240.1 (8) K. There is additionally a higher temperature region with \textit{T}$_{JT}$ = 457.8(10) K which may be related to the onset of JT distortion of the octahedra and the presence of short-mid range cooperative motions (Fig. \ \ref{fig:s8}). Details of the fitting are provided in the SI \cite{Crawford2025SignaturesMaterial}. The merit in this analysis is that the strain can be determined very precisely when using data approaching a resolution of 10$^{-5}$ (i.e. the high resolution ID22 data), whereas distortion modes typically have a high degree of uncertainty associated with them on account of correlations and systematic errors.

By looking at the ordering of individual bond lengths within the structure (Fig.\ \ref{fig4}(d)), there are alternating short and long Fe-F bonds perpendicular to \textit{b}, and a second long bond is aligned along \textit{b}. This results in a compressed octahedron with an alternating compression along the [101] and [10-1] directions between neighbouring octahedra in the \textit{ac} plane. This occurs due to the 90° rotation of the symmetry equivalent, corner-sharing octahedra, that may be assigned to have an alternation of the doubly occupied \textit{t$_{2g}$} orbital: the \textit{d$_{xz}$} and \textit{d$_{yz}$}.  Since there is no alternating of bond lengths along \textit{b} and rather ‘columns’ of octahedra elongated along one direction (\textit{b}) occur, the OO pattern is described as C-type OO. Fig.\ \ref{fig4}(d) depicts the alternating ordering of the \textit{d$_{xz}$} and \textit{d$_{yz}$} orbitals through a schematic representation.

\begin{figure}
\includegraphics[scale=1]{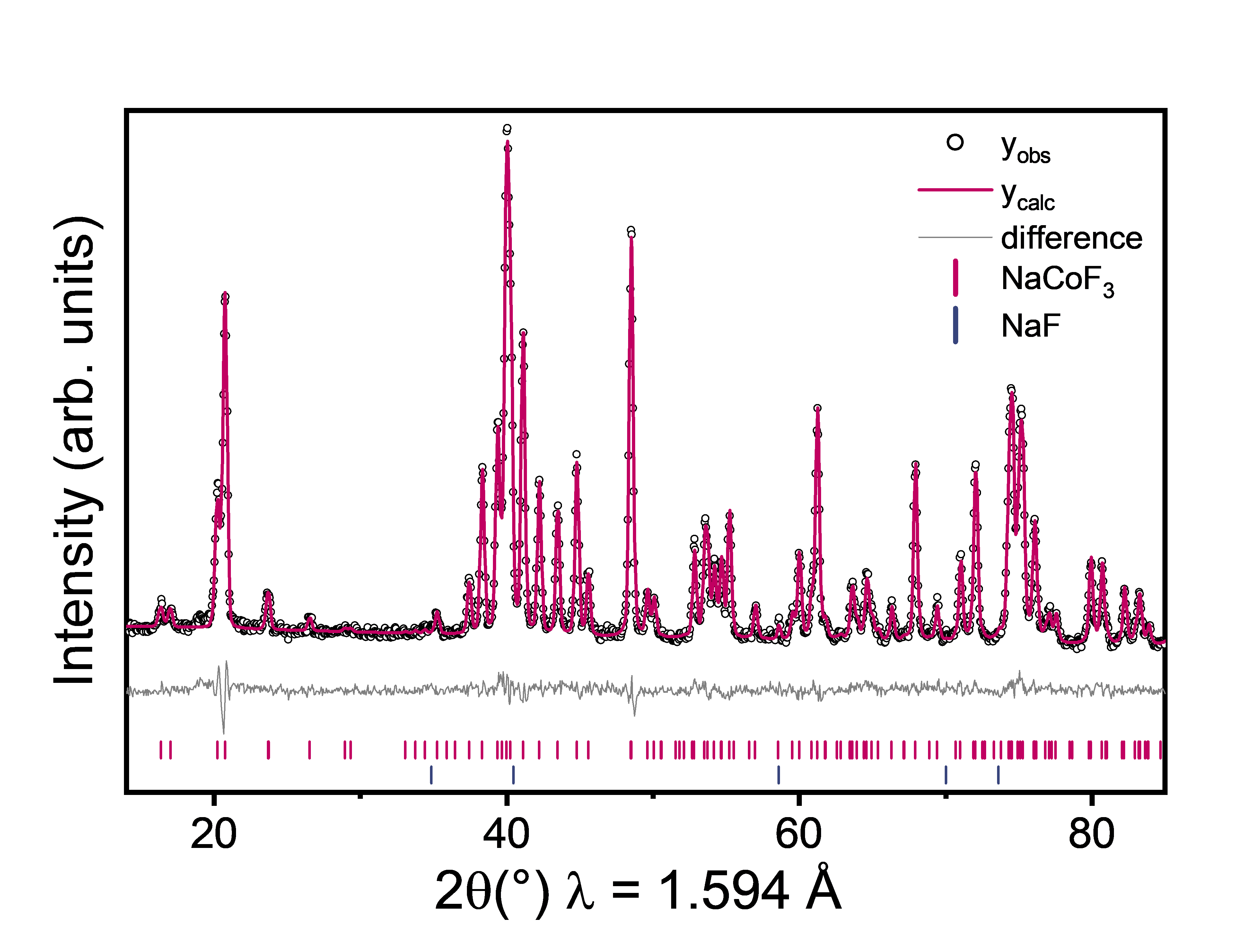}
\caption{\label{fig5}\textbf{a}. Rietveld refinement of NaCoF$_3$ in \textit{Pnma}.1 (BNS number 62.441) using neutron diffraction data collected at D2B ($\lambda$ = 1.594 Å) at 4 K. An NaF impurity is present corresponding to a phase fraction of 0.37 \%. }
\end{figure}

The absence of any deviation in \textit{M} = Co from the general trend in $\Gamma_3^+$ observed, allows us to rule out the presence of any C-type orbital ordering in this compound down to 4 K. Additionally no monoclinic distortions or supercell reflections were observed in the diffraction data, which would have indicated the presence of any symmetry lowering associated with the doubling of a unit cell axis. Given that NaCoF$_3$ is an AFM insulator with a fully saturated magnetic moment of 2.89(2) $\mu_B$/Co, this implies it has an orbitally-disordered ground state in which any orbital angular momentum is fully quenched, rather than a high spin - low spin (\textit{t$_{2g}$}$^6$) transition or adopting a fully delocalised metallic state. (See magnetic susceptibility indicating high spin Co$^{2+}$, and magnetic structure in Fig.\ \ref{fig:s9}(b,d) \cite{Crawford2025SignaturesMaterial} determined by refinement against neutron diffraction data (Fig.\ \ref{fig5})). It is interesting to compare NaCoF$_3$ with the LnVO$_3$ series since one might expect them to have strong similarities in the JT distortions associated with \textit{t$_{2g}$}$^5$ and \textit{t$_{2g}$}$^2$ degeneracies and hence the nature of the orbital order. As stated in the introduction, LnVO$_3$ have G-type OO transitions, with the smaller Ln hosting a further lower temperature reentrant C-type OO transiton (\textit{Pnma}) \cite{Shanbhag2023ObservationLaVO3}\cite{Sage2007CompetingExpansion}\cite{Varignon2015CouplingPerovskites}. In YVO$_3$ the C-type OO phase transition temperature to \textit{Pnma} symmetry from one that has G-type orbital order \cite{Blake2002NeutronYVO3} below approximately 60 K, and above 220 K the G-type orbital order disappears. In comparison NaCoF$_3$ shows no evidence of symmetry lowering to a G-type OO phase or ordering of bond lengths to a C-type OO in \textit{Pnma}.

Since the tolerance factor for both YVO$_3$ ($\tau$ =  0.836) and NaCoF$_3$ ($\tau$ =  0.842) are similar and their structures would be assumed to have similar degrees of octahedral tilting it may be surprising their apparent discrepancy between the adopted ground states. Comparing the structures at room temperature, above any orbital ordering, we find YVO$_3$ has negative $\Gamma_3^+$ (-0.016) indicating a tetragonal compression. NaCoF$_3$ on the other hand has a value that is much closer to zero (~0.001). This intrinsic strain biases the LnVO$_3$ to form G-type OO below the ordering temperature \cite{Blake2002NeutronYVO3} and hence offers a reason why a similar form of OO is absent in NaCoF$_3$.

\section{\label{sec:level1}Conclusions}
Our studies have shown the presence of \textit{t$_{2g}$} orbital order in fluoride perovskites by analysis of the octahedral distortions. We show that Jahn-Teller distortions are present in NaFeF$_3$, and manifest as an ordering of a fully occupied \textit{d$_{xz}$} and \textit{d$_{yz}$} orbital in a C-type ordered fashion as allowed in \textit{Pnma}. 

We further show that NaFeF$_3$, undergoes a C-type orbital order transition that is driven by the electronic degeneracy of the local \textit{t$_{2g}^4$} states of Fe$^{2+}$, producing a Jahn-Teller distortion with 2 short, 4 long bonds in accordance with expectations based on crystal field arguments. Surprisingly, the apparently isosymmetric phase transition is second-order like, inconsistent with thermodynamic requirements and thus imply there must be an associated hidden symmetry breaking. On the other hand, the careful consideration of the temperature evolution of the symmetry adapted strain in NaCoF$_3$ allows us to confidently rule out the occurrence of any long range orbital orders. Since NaCoF$_3$ is an insulator in which orbital angular momentum is evidently fully quenched below \textit{T}$_N$, our findings point towards a novel orbital disorder associated to the \textit{t$_{2g}^5$} electronic degeneracy. Our findings should motivate further investigations into the factors that can control and tune the cross-over from orbital order to disorder, with implications for technologically relevant properties.

\begin{acknowledgments}
We acknowledge the European Synchrotron Radiation Facility (ESRF) for the provision of synchrotron
facilities under proposal number HC5601 (doi:10.15151/ESRF-ES-1467550327). We thank the Institut Laue-Langevin for providing neutron facilities under experiment number 5-24-737 (doi:10.5291/ILL-DATA.5-24-737). We thank J. Gainza-Martin, A. N. Fitch
and C. Dejoie for support in using ID22 at the ESRF and C. Ritter for support in neutron experiments at the ILL. We additionally thank Diamond Light Source under the OxfordWarwick Solid-State Chemistry BAG (CY32893-10, CY32893-8). We would like to thank P. Sinted for support on experiment beamtime at ESRF. Initial sample characterization was performed using equipment provided by the Warwick X-ray Research Technology Platforms and we thank M. R. Lees for use of SQUID magnetometers to perform magnetometry measurements at the Department of Physics, University of Warwick. R.I.W. and M.S.S. acknowledge the Leverhulme Trust for a research project grant (Grant No. RPG-2022-22). M.S.S acknowledges funding from the Royal Society (UF160265 and URFR231012). 
\end{acknowledgments}

\nocite{*}
\bibliography{References}

\providecommand{\noopsort}[1]{}\providecommand{\singleletter}[1]{#1}%
\begin{thebibliography}{39}%
\makeatletter
\providecommand \@ifxundefined [1]{%
 \@ifx{#1\undefined}
}%
\providecommand \@ifnum [1]{%
 \ifnum #1\expandafter \@firstoftwo
 \else \expandafter \@secondoftwo
 \fi
}%
\providecommand \@ifx [1]{%
 \ifx #1\expandafter \@firstoftwo
 \else \expandafter \@secondoftwo
 \fi
}%
\providecommand \natexlab [1]{#1}%
\providecommand \enquote  [1]{``#1''}%
\providecommand \bibnamefont  [1]{#1}%
\providecommand \bibfnamefont [1]{#1}%
\providecommand \citenamefont [1]{#1}%
\providecommand \href@noop [0]{\@secondoftwo}%
\providecommand \href [0]{\begingroup \@sanitize@url \@href}%
\providecommand \@href[1]{\@@startlink{#1}\@@href}%
\providecommand \@@href[1]{\endgroup#1\@@endlink}%
\providecommand \@sanitize@url [0]{\catcode `\\12\catcode `\$12\catcode
  `\&12\catcode `\#12\catcode `\^12\catcode `\_12\catcode `\%12\relax}%
\providecommand \@@startlink[1]{}%
\providecommand \@@endlink[0]{}%
\providecommand \url  [0]{\begingroup\@sanitize@url \@url }%
\providecommand \@url [1]{\endgroup\@href {#1}{\urlprefix }}%
\providecommand \urlprefix  [0]{URL }%
\providecommand \Eprint [0]{\href }%
\providecommand \doibase [0]{https://doi.org/}%
\providecommand \selectlanguage [0]{\@gobble}%
\providecommand \bibinfo  [0]{\@secondoftwo}%
\providecommand \bibfield  [0]{\@secondoftwo}%
\providecommand \translation [1]{[#1]}%
\providecommand \BibitemOpen [0]{}%
\providecommand \bibitemStop [0]{}%
\providecommand \bibitemNoStop [0]{.\EOS\space}%
\providecommand \EOS [0]{\spacefactor3000\relax}%
\providecommand \BibitemShut  [1]{\csname bibitem#1\endcsname}%
\let\auto@bib@innerbib\@empty
\bibitem [{\citenamefont {Mitchell}(2002)}]{Mitchell2002Perovskites:Ancient}%
  \BibitemOpen
  \bibfield  {author} {\bibinfo {author} {\bibfnamefont {R.~H.}\ \bibnamefont
  {Mitchell}},\ }\href
  {https://www.for.gov.bc.ca/hts/risc/pubs/tebiodiv/snakes/assets/snake.pdf}
  {\emph {\bibinfo {title} {{Perovskites: modern and ancient}}}},\ \bibinfo
  {number} {38}\ (\bibinfo  {publisher} {Almaz Press Inc.},\ \bibinfo {address}
  {Thunder Bay, Ontario},\ \bibinfo {year} {2002})\BibitemShut {NoStop}%
\bibitem [{\citenamefont {Lufaso}\ and\ \citenamefont
  {Woodward}(2004)}]{Lufaso2004Jahn-TellerPerovskites}%
  \BibitemOpen
  \bibfield  {author} {\bibinfo {author} {\bibfnamefont {M.~W.}\ \bibnamefont
  {Lufaso}}\ and\ \bibinfo {author} {\bibfnamefont {P.~M.}\ \bibnamefont
  {Woodward}},\ }\bibfield  {title} {\bibinfo {title} {{Jahn-Teller
  distortions, cation ordering and octahedral tilting in perovskites}},\ }\href
  {https://doi.org/10.1107/S0108768103026661} {\bibfield  {journal} {\bibinfo
  {journal} {Acta Crystallogr. B}\ }\textbf {\bibinfo {volume} {60}},\ \bibinfo
  {pages} {10} (\bibinfo {year} {2004})}\BibitemShut {NoStop}%
\bibitem [{\citenamefont {Streltsov}\ and\ \citenamefont
  {Khomskii}(2017)}]{Streltsov2017OrbitalTrends}%
  \BibitemOpen
  \bibfield  {author} {\bibinfo {author} {\bibfnamefont {S.~V.}\ \bibnamefont
  {Streltsov}}\ and\ \bibinfo {author} {\bibfnamefont {D.~I.}\ \bibnamefont
  {Khomskii}},\ }\bibfield  {title} {\bibinfo {title} {{Orbital physics in
  transition metal compounds: new trends}},\ }\href
  {https://doi.org/10.3367/ufnr.2017.08.038196} {\bibfield  {journal} {\bibinfo
   {journal} {Physics-Uspekhi}\ }\textbf {\bibinfo {volume} {187}},\ \bibinfo
  {pages} {1205} (\bibinfo {year} {2017})}\BibitemShut {NoStop}%
\bibitem [{\citenamefont {Goodenough}\ and\ \citenamefont
  {Zhou}(2007)}]{Goodenough2007OrbitalPerovskites}%
  \BibitemOpen
  \bibfield  {author} {\bibinfo {author} {\bibfnamefont {J.~B.}\ \bibnamefont
  {Goodenough}}\ and\ \bibinfo {author} {\bibfnamefont {J.~S.}\ \bibnamefont
  {Zhou}},\ }\bibfield  {title} {\bibinfo {title} {{Orbital ordering in
  orthorhombic perovskites}},\ }\href {https://doi.org/10.1039/b701805c}
  {\bibfield  {journal} {\bibinfo  {journal} {Journal of Materials Chemistry}\
  }\textbf {\bibinfo {volume} {17}},\ \bibinfo {pages} {2394} (\bibinfo {year}
  {2007})}\BibitemShut {NoStop}%
\bibitem [{\citenamefont {Bernal}\ \emph
  {et~al.}(2020{\natexlab{a}})\citenamefont {Bernal}, \citenamefont {Sottmann},
  \citenamefont {Wragg}, \citenamefont {Fjellv{\aa}g}, \citenamefont
  {Fjellv{\aa}g}, \citenamefont {Drathen}, \citenamefont {S{\l}awi{\'{n}}ski},\
  and\ \citenamefont {L{\o}vvik}}]{Bernal2020StructuralNaCrF3}%
  \BibitemOpen
  \bibfield  {author} {\bibinfo {author} {\bibfnamefont {F.~L.~M.}\
  \bibnamefont {Bernal}}, \bibinfo {author} {\bibfnamefont {J.}~\bibnamefont
  {Sottmann}}, \bibinfo {author} {\bibfnamefont {D.~S.}\ \bibnamefont {Wragg}},
  \bibinfo {author} {\bibfnamefont {H.}~\bibnamefont {Fjellv{\aa}g}}, \bibinfo
  {author} {\bibfnamefont {O.~S.}\ \bibnamefont {Fjellv{\aa}g}}, \bibinfo
  {author} {\bibfnamefont {C.}~\bibnamefont {Drathen}}, \bibinfo {author}
  {\bibfnamefont {W.~A.}\ \bibnamefont {S{\l}awi{\'{n}}ski}},\ and\ \bibinfo
  {author} {\bibfnamefont {O.~M.}\ \bibnamefont {L{\o}vvik}},\ }\bibfield
  {title} {\bibinfo {title} {{Structural and magnetic characterization of the
  elusive Jahn-Teller active NaCrF$_3$}},\ }\href
  {https://doi.org/10.1103/PhysRevMaterials.4.054412} {\bibfield  {journal}
  {\bibinfo  {journal} {Phys. Rev. Mater.}\ }\textbf {\bibinfo {volume} {4}},\
  \bibinfo {pages} {054412} (\bibinfo {year} {2020}{\natexlab{a}})}\BibitemShut
  {NoStop}%
\bibitem [{\citenamefont {Zhou}\ \emph {et~al.}(2011)\citenamefont {Zhou},
  \citenamefont {Alonso}, \citenamefont {Han}, \citenamefont
  {Fern{\'{a}}ndez-D{\'{i}}az}, \citenamefont {Cheng},\ and\ \citenamefont
  {Goodenough}}]{Zhou2011Jahn-TellerPressure}%
  \BibitemOpen
  \bibfield  {author} {\bibinfo {author} {\bibfnamefont {J.~S.}\ \bibnamefont
  {Zhou}}, \bibinfo {author} {\bibfnamefont {J.~A.}\ \bibnamefont {Alonso}},
  \bibinfo {author} {\bibfnamefont {J.~T.}\ \bibnamefont {Han}}, \bibinfo
  {author} {\bibfnamefont {M.~T.}\ \bibnamefont {Fern{\'{a}}ndez-D{\'{i}}az}},
  \bibinfo {author} {\bibfnamefont {J.~G.}\ \bibnamefont {Cheng}},\ and\
  \bibinfo {author} {\bibfnamefont {J.~B.}\ \bibnamefont {Goodenough}},\
  }\bibfield  {title} {\bibinfo {title} {{Jahn-Teller distortion in perovskite
  KCuF$_3$ under high pressure}},\ }\href
  {https://doi.org/10.1016/j.jfluchem.2011.06.047} {\bibfield  {journal}
  {\bibinfo  {journal} {J. Fluorine Chem.}\ }\textbf {\bibinfo {volume}
  {132}},\ \bibinfo {pages} {1117} (\bibinfo {year} {2011})}\BibitemShut
  {NoStop}%
\bibitem [{\citenamefont {Tragheim}\ \emph {et~al.}(2024)\citenamefont
  {Tragheim}, \citenamefont {Ritter},\ and\ \citenamefont
  {Senn}}]{Tragheim2024ProximityPerovskites}%
  \BibitemOpen
  \bibfield  {author} {\bibinfo {author} {\bibfnamefont {B.~R.}\ \bibnamefont
  {Tragheim}}, \bibinfo {author} {\bibfnamefont {C.}~\bibnamefont {Ritter}},\
  and\ \bibinfo {author} {\bibfnamefont {M.~S.}\ \bibnamefont {Senn}},\
  }\bibfield  {title} {\bibinfo {title} {{Proximity to a state with orbital
  order and charge disorder in optimally doped R$_{5/8}$Ca$_{3/8}$MnO$_3$
  perovskites}},\ }\href {https://doi.org/10.1103/PhysRevB.110.235141}
  {\bibfield  {journal} {\bibinfo  {journal} {Phys. Rev. B}\ }\textbf {\bibinfo
  {volume} {110}},\ \bibinfo {pages} {235141} (\bibinfo {year}
  {2024})}\BibitemShut {NoStop}%
\bibitem [{\citenamefont {Pissas}\ \emph {et~al.}(2005)\citenamefont {Pissas},
  \citenamefont {Margiolaki}, \citenamefont {Papavassiliou}, \citenamefont
  {Stamopoulos},\ and\ \citenamefont {Argyriou}}]{Pissas2005Crystal0.11x0.175}%
  \BibitemOpen
  \bibfield  {author} {\bibinfo {author} {\bibfnamefont {M.}~\bibnamefont
  {Pissas}}, \bibinfo {author} {\bibfnamefont {I.}~\bibnamefont {Margiolaki}},
  \bibinfo {author} {\bibfnamefont {G.}~\bibnamefont {Papavassiliou}}, \bibinfo
  {author} {\bibfnamefont {D.}~\bibnamefont {Stamopoulos}},\ and\ \bibinfo
  {author} {\bibfnamefont {D.}~\bibnamefont {Argyriou}},\ }\bibfield  {title}
  {\bibinfo {title} {{Crystal and magnetic structure of the
  La$_{1-x}$Ca$_x$MnO$_3$ compound (0.11=x=0.175)}},\ }\href
  {https://doi.org/10.1103/PhysRevB.72.064425} {\bibfield  {journal} {\bibinfo
  {journal} {Phys. Rev. B}\ }\textbf {\bibinfo {volume} {72}},\ \bibinfo
  {pages} {064425} (\bibinfo {year} {2005})}\BibitemShut {NoStop}%
\bibitem [{\citenamefont {Zhang}\ \emph {et~al.}(2020)\citenamefont {Zhang},
  \citenamefont {Koch},\ and\ \citenamefont
  {Pavarini}}]{Zhang2020OriginLaTiO3}%
  \BibitemOpen
  \bibfield  {author} {\bibinfo {author} {\bibfnamefont {X.~J.}\ \bibnamefont
  {Zhang}}, \bibinfo {author} {\bibfnamefont {E.}~\bibnamefont {Koch}},\ and\
  \bibinfo {author} {\bibfnamefont {E.}~\bibnamefont {Pavarini}},\ }\bibfield
  {title} {\bibinfo {title} {{Origin of orbital ordering in YTiO$_3$ and
  LaTiO$_3$}},\ }\href {https://doi.org/10.1103/PhysRevB.102.035113} {\bibfield
   {journal} {\bibinfo  {journal} {Phys. Rev. B}\ }\textbf {\bibinfo {volume}
  {102}},\ \bibinfo {pages} {035113} (\bibinfo {year} {2020})}\BibitemShut
  {NoStop}%
\bibitem [{\citenamefont {Varignon}\ \emph {et~al.}(2015)\citenamefont
  {Varignon}, \citenamefont {Bristowe}, \citenamefont {Bousquet},\ and\
  \citenamefont {Ghosez}}]{Varignon2015CouplingPerovskites}%
  \BibitemOpen
  \bibfield  {author} {\bibinfo {author} {\bibfnamefont {J.}~\bibnamefont
  {Varignon}}, \bibinfo {author} {\bibfnamefont {N.~C.}\ \bibnamefont
  {Bristowe}}, \bibinfo {author} {\bibfnamefont {E.}~\bibnamefont {Bousquet}},\
  and\ \bibinfo {author} {\bibfnamefont {P.}~\bibnamefont {Ghosez}},\
  }\bibfield  {title} {\bibinfo {title} {{Coupling and electrical control of
  structural, orbital and magnetic orders in perovskites}},\ }\href
  {https://doi.org/10.1038/srep15364} {\bibfield  {journal} {\bibinfo
  {journal} {Sci. Rep.}\ }\textbf {\bibinfo {volume} {5}},\ \bibinfo {pages}
  {15364} (\bibinfo {year} {2015})}\BibitemShut {NoStop}%
\bibitem [{\citenamefont {Johnson}\ \emph {et~al.}(2012)\citenamefont
  {Johnson}, \citenamefont {Tang}, \citenamefont {Evans}, \citenamefont
  {Bland}, \citenamefont {Free}, \citenamefont {Beale}, \citenamefont {Hatton},
  \citenamefont {Bouchenoire}, \citenamefont {Prabhakaran},\ and\ \citenamefont
  {Boothroyd}}]{Johnson2012X-raySmVO3}%
  \BibitemOpen
  \bibfield  {author} {\bibinfo {author} {\bibfnamefont {R.~D.}\ \bibnamefont
  {Johnson}}, \bibinfo {author} {\bibfnamefont {C.~C.}\ \bibnamefont {Tang}},
  \bibinfo {author} {\bibfnamefont {I.~R.}\ \bibnamefont {Evans}}, \bibinfo
  {author} {\bibfnamefont {S.~R.}\ \bibnamefont {Bland}}, \bibinfo {author}
  {\bibfnamefont {D.~G.}\ \bibnamefont {Free}}, \bibinfo {author}
  {\bibfnamefont {T.~A.}\ \bibnamefont {Beale}}, \bibinfo {author}
  {\bibfnamefont {P.~D.}\ \bibnamefont {Hatton}}, \bibinfo {author}
  {\bibfnamefont {L.}~\bibnamefont {Bouchenoire}}, \bibinfo {author}
  {\bibfnamefont {D.}~\bibnamefont {Prabhakaran}},\ and\ \bibinfo {author}
  {\bibfnamefont {A.~T.}\ \bibnamefont {Boothroyd}},\ }\bibfield  {title}
  {\bibinfo {title} {{X-ray diffraction study of the temperature-induced
  structural phase transitions in SmVO$_3$}},\ }\href
  {https://doi.org/10.1103/PhysRevB.85.224102} {\bibfield  {journal} {\bibinfo
  {journal} {Phys. Rev. B}\ }\textbf {\bibinfo {volume} {85}},\ \bibinfo
  {pages} {224102} (\bibinfo {year} {2012})}\BibitemShut {NoStop}%
\bibitem [{\citenamefont {Sage}\ \emph {et~al.}(2007)\citenamefont {Sage},
  \citenamefont {Blake}, \citenamefont {Marquina},\ and\ \citenamefont
  {Palstra}}]{Sage2007CompetingExpansion}%
  \BibitemOpen
  \bibfield  {author} {\bibinfo {author} {\bibfnamefont {M.~H.}\ \bibnamefont
  {Sage}}, \bibinfo {author} {\bibfnamefont {G.~R.}\ \bibnamefont {Blake}},
  \bibinfo {author} {\bibfnamefont {C.}~\bibnamefont {Marquina}},\ and\
  \bibinfo {author} {\bibfnamefont {T.~T.}\ \bibnamefont {Palstra}},\
  }\bibfield  {title} {\bibinfo {title} {{Competing orbital ordering in RVO$_3$
  compounds: High-resolution x-ray diffraction and thermal expansion}},\ }\href
  {https://doi.org/10.1103/PhysRevB.76.195102} {\bibfield  {journal} {\bibinfo
  {journal} {Phys. Rev. B}\ }\textbf {\bibinfo {volume} {76}},\ \bibinfo
  {pages} {195102} (\bibinfo {year} {2007})}\BibitemShut {NoStop}%
\bibitem [{\citenamefont {Ritter}\ \emph {et~al.}(2016)\citenamefont {Ritter},
  \citenamefont {Ivanov}, \citenamefont {Bazuev},\ and\ \citenamefont
  {Fauth}}]{Ritter2016CrystallographicTmVO3}%
  \BibitemOpen
  \bibfield  {author} {\bibinfo {author} {\bibfnamefont {C.}~\bibnamefont
  {Ritter}}, \bibinfo {author} {\bibfnamefont {S.~A.}\ \bibnamefont {Ivanov}},
  \bibinfo {author} {\bibfnamefont {G.~V.}\ \bibnamefont {Bazuev}},\ and\
  \bibinfo {author} {\bibfnamefont {F.}~\bibnamefont {Fauth}},\ }\bibfield
  {title} {\bibinfo {title} {{Crystallographic phase coexistence, spin-orbital
  order transitions, and spontaneous spin flop in TmVO$_3$}},\ }\href
  {https://doi.org/10.1103/PhysRevB.93.054423} {\bibfield  {journal} {\bibinfo
  {journal} {Phys. Rev. B}\ }\textbf {\bibinfo {volume} {93}},\ \bibinfo
  {pages} {054423} (\bibinfo {year} {2016})}\BibitemShut {NoStop}%
\bibitem [{\citenamefont {Bull}\ and\ \citenamefont
  {Knight}(2016)}]{Bull2016Low-temperatureStudy}%
  \BibitemOpen
  \bibfield  {author} {\bibinfo {author} {\bibfnamefont {C.~L.}\ \bibnamefont
  {Bull}}\ and\ \bibinfo {author} {\bibfnamefont {K.~S.}\ \bibnamefont
  {Knight}},\ }\bibfield  {title} {\bibinfo {title} {{Low-temperature
  structural behaviour of LaCoO$_3$ - A high-resolution neutron study}},\
  }\href {https://doi.org/10.1016/j.solidstatesciences.2016.04.012} {\bibfield
  {journal} {\bibinfo  {journal} {Solid State Sci.}\ }\textbf {\bibinfo
  {volume} {57}},\ \bibinfo {pages} {38} (\bibinfo {year} {2016})}\BibitemShut
  {NoStop}%
\bibitem [{\citenamefont {Amit}\ \emph {et~al.}(1974)\citenamefont {Amit},
  \citenamefont {Horowitz},\ and\ \citenamefont
  {Makovsky}}]{Amit1974PreparationKFeBr3}%
  \BibitemOpen
  \bibfield  {author} {\bibinfo {author} {\bibfnamefont {M.}~\bibnamefont
  {Amit}}, \bibinfo {author} {\bibfnamefont {A.}~\bibnamefont {Horowitz}},\
  and\ \bibinfo {author} {\bibfnamefont {J.}~\bibnamefont {Makovsky}},\
  }\bibfield  {title} {\bibinfo {title} {{Preparation and Crystal Structure of
  KFeCl$_3$ and KFeBr$_3$}},\ }\href {https://doi.org/10.1002/ijch.197400071}
  {\bibfield  {journal} {\bibinfo  {journal} {Isr. J. Chem.}\ }\textbf
  {\bibinfo {volume} {12}},\ \bibinfo {pages} {827} (\bibinfo {year}
  {1974})}\BibitemShut {NoStop}%
\bibitem [{\citenamefont {Kohne}\ \emph {et~al.}(1993)\citenamefont {Kohne},
  \citenamefont {Kemnitz}, \citenamefont {Mattausch},\ and\ \citenamefont
  {Simon}}]{Kohne1993CrystalCsFeC3}%
  \BibitemOpen
  \bibfield  {author} {\bibinfo {author} {\bibfnamefont {A.}~\bibnamefont
  {Kohne}}, \bibinfo {author} {\bibfnamefont {E.}~\bibnamefont {Kemnitz}},
  \bibinfo {author} {\bibfnamefont {H.}~\bibnamefont {Mattausch}},\ and\
  \bibinfo {author} {\bibfnamefont {A.}~\bibnamefont {Simon}},\ }\bibfield
  {title} {\bibinfo {title} {{Crystal structure of caesium
  trichloroferrate(II), CsFeC$_3$}},\ }\href
  {https://doi.org/10.1524/zkri.1993.203.12.316} {\bibfield  {journal}
  {\bibinfo  {journal} {Z. Kristallogr. Cryst. Mater.}\ }\textbf {\bibinfo
  {volume} {203}},\ \bibinfo {pages} {316} (\bibinfo {year}
  {1993})}\BibitemShut {NoStop}%
\bibitem [{\citenamefont {Takeda}\ \emph {et~al.}(1974)\citenamefont {Takeda},
  \citenamefont {Shimada}, \citenamefont {Kanamaru},\ and\ \citenamefont
  {Koizumi}}]{Takeda1974StructureCsFeBr3}%
  \BibitemOpen
  \bibfield  {author} {\bibinfo {author} {\bibfnamefont {Y.}~\bibnamefont
  {Takeda}}, \bibinfo {author} {\bibfnamefont {M.}~\bibnamefont {Shimada}},
  \bibinfo {author} {\bibfnamefont {F.}~\bibnamefont {Kanamaru}},\ and\
  \bibinfo {author} {\bibfnamefont {M.}~\bibnamefont {Koizumi}},\ }\bibfield
  {title} {\bibinfo {title} {{Structure and Properties of CsFeBr$_3$}},\ }\href
  {https://doi.org/10.1143/JPSJ.37.276} {\bibfield  {journal} {\bibinfo
  {journal} {J. Phys. Soc. Jpn.}\ }\textbf {\bibinfo {volume} {37}},\ \bibinfo
  {pages} {276} (\bibinfo {year} {1974})}\BibitemShut {NoStop}%
\bibitem [{\citenamefont {Bernal}\ \emph
  {et~al.}(2020{\natexlab{b}})\citenamefont {Bernal}, \citenamefont {Gonano},
  \citenamefont {Lundvall}, \citenamefont {Wragg}, \citenamefont
  {Fjellv{\aa}g}, \citenamefont {Veillon}, \citenamefont {S{\l}awi{\'{n}}ski},\
  and\ \citenamefont {Fjellvag}}]{Bernal2020CantedMethod}%
  \BibitemOpen
  \bibfield  {author} {\bibinfo {author} {\bibfnamefont {F.~L.~M.}\
  \bibnamefont {Bernal}}, \bibinfo {author} {\bibfnamefont {B.}~\bibnamefont
  {Gonano}}, \bibinfo {author} {\bibfnamefont {F.}~\bibnamefont {Lundvall}},
  \bibinfo {author} {\bibfnamefont {D.~S.}\ \bibnamefont {Wragg}}, \bibinfo
  {author} {\bibfnamefont {H.}~\bibnamefont {Fjellv{\aa}g}}, \bibinfo {author}
  {\bibfnamefont {F.}~\bibnamefont {Veillon}}, \bibinfo {author} {\bibfnamefont
  {W.~A.}\ \bibnamefont {S{\l}awi{\'{n}}ski}},\ and\ \bibinfo {author}
  {\bibfnamefont {O.~S.}\ \bibnamefont {Fjellvag}},\ }\bibfield  {title}
  {\bibinfo {title} {{Canted antiferromagnetism in high-purity NaFeF$_3$
  prepared by a novel wet-chemical synthesis method}},\ }\href
  {https://doi.org/10.1103/PhysRevMaterials.4.114412} {\bibfield  {journal}
  {\bibinfo  {journal} {Phys. Rev. Mater.}\ }\textbf {\bibinfo {volume} {4}},\
  \bibinfo {pages} {114412} (\bibinfo {year} {2020}{\natexlab{b}})}\BibitemShut
  {NoStop}%
\bibitem [{\citenamefont {Fitch}\ \emph {et~al.}(2023)\citenamefont {Fitch},
  \citenamefont {Dejoie}, \citenamefont {Covacci}, \citenamefont
  {Confalonieri}, \citenamefont {Grendal}, \citenamefont {Claustre},
  \citenamefont {Guillou}, \citenamefont {Kieffer}, \citenamefont {De~Nolf},
  \citenamefont {Petitdemange}, \citenamefont {Ruat},\ and\ \citenamefont
  {Watier}}]{Fitch2023ID22ESRF}%
  \BibitemOpen
  \bibfield  {author} {\bibinfo {author} {\bibfnamefont {A.}~\bibnamefont
  {Fitch}}, \bibinfo {author} {\bibfnamefont {C.}~\bibnamefont {Dejoie}},
  \bibinfo {author} {\bibfnamefont {E.}~\bibnamefont {Covacci}}, \bibinfo
  {author} {\bibfnamefont {G.}~\bibnamefont {Confalonieri}}, \bibinfo {author}
  {\bibfnamefont {O.}~\bibnamefont {Grendal}}, \bibinfo {author} {\bibfnamefont
  {L.}~\bibnamefont {Claustre}}, \bibinfo {author} {\bibfnamefont
  {P.}~\bibnamefont {Guillou}}, \bibinfo {author} {\bibfnamefont
  {J.}~\bibnamefont {Kieffer}}, \bibinfo {author} {\bibfnamefont
  {W.}~\bibnamefont {De~Nolf}}, \bibinfo {author} {\bibfnamefont
  {S.}~\bibnamefont {Petitdemange}}, \bibinfo {author} {\bibfnamefont
  {M.}~\bibnamefont {Ruat}},\ and\ \bibinfo {author} {\bibfnamefont
  {Y.}~\bibnamefont {Watier}},\ }\bibfield  {title} {\bibinfo {title} {{ID22 -
  the high-resolution powder-diffraction beamline at ESRF}},\ }\href
  {https://doi.org/10.1107/S1600577523004915} {\bibfield  {journal} {\bibinfo
  {journal} {J. Synchrotron Radiat.}\ }\textbf {\bibinfo {volume} {30}},\
  \bibinfo {pages} {1003} (\bibinfo {year} {2023})}\BibitemShut {NoStop}%
\bibitem [{\citenamefont {Coelho}(2017)}]{Coelho2017TOPAS-Academic}%
  \BibitemOpen
  \bibfield  {author} {\bibinfo {author} {\bibfnamefont {A.~A.}\ \bibnamefont
  {Coelho}},\ }\href {www.topas-academic.net} {\bibinfo {title}
  {{TOPAS-Academic}}} (\bibinfo {year} {2017})\BibitemShut {NoStop}%
\bibitem [{\citenamefont {Campbell}\ \emph {et~al.}(2006)\citenamefont
  {Campbell}, \citenamefont {Stokes}, \citenamefont {Tanner},\ and\
  \citenamefont {Hatch}}]{Campbell2006ISODISPLACE:Distortions}%
  \BibitemOpen
  \bibfield  {author} {\bibinfo {author} {\bibfnamefont {B.~J.}\ \bibnamefont
  {Campbell}}, \bibinfo {author} {\bibfnamefont {H.~T.}\ \bibnamefont
  {Stokes}}, \bibinfo {author} {\bibfnamefont {D.~E.}\ \bibnamefont {Tanner}},\
  and\ \bibinfo {author} {\bibfnamefont {D.~M.}\ \bibnamefont {Hatch}},\
  }\bibfield  {title} {\bibinfo {title} {{ISODISPLACE: A web-based tool for
  exploring structural distortions}},\ }\href
  {https://doi.org/10.1107/S0021889806014075} {\bibfield  {journal} {\bibinfo
  {journal} {J. Appl. Crystallogr.}\ }\textbf {\bibinfo {volume} {39}},\
  \bibinfo {pages} {607} (\bibinfo {year} {2006})}\BibitemShut {NoStop}%
\bibitem [{\citenamefont {Stokes}\ \emph {et~al.}({\natexlab{a}})\citenamefont
  {Stokes}, \citenamefont {Hatch},\ and\ \citenamefont
  {Campbell}}]{StokesISODISTORTSuite}%
  \BibitemOpen
  \bibfield  {author} {\bibinfo {author} {\bibfnamefont {H.~T.}\ \bibnamefont
  {Stokes}}, \bibinfo {author} {\bibfnamefont {D.~M.}\ \bibnamefont {Hatch}},\
  and\ \bibinfo {author} {\bibfnamefont {B.~J.}\ \bibnamefont {Campbell}},\
  }\href {iso.byu.edu} {\bibinfo {title} {{ISODISTORT, ISOTROPY Software
  Suite}}} ({\natexlab{a}})\BibitemShut {NoStop}%
\bibitem [{\citenamefont {Stokes}\ \emph {et~al.}({\natexlab{b}})\citenamefont
  {Stokes}, \citenamefont {Hatch},\ and\ \citenamefont
  {Campbell}}]{StokesINVARIANTSIso.byu.edu.}%
  \BibitemOpen
  \bibfield  {author} {\bibinfo {author} {\bibfnamefont {H.~T.}\ \bibnamefont
  {Stokes}}, \bibinfo {author} {\bibfnamefont {D.~M.}\ \bibnamefont {Hatch}},\
  and\ \bibinfo {author} {\bibfnamefont {B.~J.}\ \bibnamefont {Campbell}},\
  }\href {iso.byu.edu.} {\bibinfo {title} {{INVARIANTS, ISOTROPY Software
  Suite, iso.byu.edu.}}} ({\natexlab{b}})\BibitemShut {NoStop}%
\bibitem [{\citenamefont {Hatch}\ and\ \citenamefont
  {Stokes}(2003)}]{Hatch2003INVARIANTS:Group}%
  \BibitemOpen
  \bibfield  {author} {\bibinfo {author} {\bibfnamefont {D.~M.}\ \bibnamefont
  {Hatch}}\ and\ \bibinfo {author} {\bibfnamefont {H.~T.}\ \bibnamefont
  {Stokes}},\ }\bibfield  {title} {\bibinfo {title} {{INVARIANTS: program for
  obtaining a list of invariant polynomials of the order-parameter components
  associated with irreducible representations of a space group}},\ }\href
  {https://doi.org/10.1107/S0021889803005946} {\bibfield  {journal} {\bibinfo
  {journal} {J. Appl. Crystallogr.}\ }\textbf {\bibinfo {volume} {36}},\
  \bibinfo {pages} {951} (\bibinfo {year} {2003})}\BibitemShut {NoStop}%
\bibitem [{\citenamefont {Momma}\ and\ \citenamefont
  {Izumi}(2011)}]{Momma2011VESTA3Data}%
  \BibitemOpen
  \bibfield  {author} {\bibinfo {author} {\bibfnamefont {K.}~\bibnamefont
  {Momma}}\ and\ \bibinfo {author} {\bibfnamefont {F.}~\bibnamefont {Izumi}},\
  }\bibfield  {title} {\bibinfo {title} {{VESTA 3 for three-dimensional
  visualization of crystal, volumetric and morphology data}},\ }\href
  {https://doi.org/10.1107/S0021889811038970} {\bibfield  {journal} {\bibinfo
  {journal} {J. Appl. Crystallogr.}\ }\textbf {\bibinfo {volume} {44}},\
  \bibinfo {pages} {1272} (\bibinfo {year} {2011})}\BibitemShut {NoStop}%
\bibitem [{\citenamefont {Nagle-Cocco}\ and\ \citenamefont
  {Dutton}(2024)}]{Nagle-Cocco2024VanVanVleckCalculator}%
  \BibitemOpen
  \bibfield  {author} {\bibinfo {author} {\bibfnamefont {L.~A.}\ \bibnamefont
  {Nagle-Cocco}}\ and\ \bibinfo {author} {\bibfnamefont {S.~E.}\ \bibnamefont
  {Dutton}},\ }\bibfield  {title} {\bibinfo {title} {{Van Vleck analysis of
  angularly distorted octahedra using VanVleckCalculator}},\ }\href
  {https://doi.org/10.1107/S1600576723009925} {\bibfield  {journal} {\bibinfo
  {journal} {J. Appl. Crystallogr.}\ }\textbf {\bibinfo {volume} {57}},\
  \bibinfo {pages} {20} (\bibinfo {year} {2024})}\BibitemShut {NoStop}%
\bibitem [{Cra()}]{Crawford2025SignaturesMaterial}%
  \BibitemOpen
  \href@noop {} {}\bibinfo {howpublished} {See Supplemental Material at [URL
  will be inserted by publisher] for variable temperature CIFs, exported
  parameters from refinements, detail on calculations of Van Vleck distortion
  modes, INVARIANTS analysis and resulting parameters, critical temperature fit
  to extract OO temperatures, magnetometry and magnetic structure analysis of
  NaCoF$_3$. The Supplemental Material also contains Refs.
  \cite{Nagle-Cocco2024VanVanVleckCalculator, Hatch2003INVARIANTS:Group,
  StokesINVARIANTSIso.byu.edu., Bernal2020CantedMethod}}\BibitemShut {NoStop}%
\bibitem [{\citenamefont {Senn}\ and\ \citenamefont
  {Bristowe}(2018)}]{Senn2018APerovskites}%
  \BibitemOpen
  \bibfield  {author} {\bibinfo {author} {\bibfnamefont {M.~S.}\ \bibnamefont
  {Senn}}\ and\ \bibinfo {author} {\bibfnamefont {N.~C.}\ \bibnamefont
  {Bristowe}},\ }\bibfield  {title} {\bibinfo {title} {{A group-theoretical
  approach to enumerating magnetoelectric and multiferroic couplings in
  perovskites}},\ }\href {https://doi.org/10.1107/S2053273318007441} {\bibfield
   {journal} {\bibinfo  {journal} {Acta Crystallogr. A}\ }\textbf {\bibinfo
  {volume} {74}},\ \bibinfo {pages} {308} (\bibinfo {year} {2018})}\BibitemShut
  {NoStop}%
\bibitem [{Note1()}]{Note1}%
  \BibitemOpen
  \bibinfo {note} {All irrep labels are given with respect to a \protect
  \textit {Pm}$\protect \bar {3}$\protect \textit {m} unit cell with the A site
  at the origin.}\BibitemShut {Stop}%
\bibitem [{\citenamefont {Van~Vleck}(1939)}]{VanVleck1939TheXY6}%
  \BibitemOpen
  \bibfield  {author} {\bibinfo {author} {\bibfnamefont {J.~H.}\ \bibnamefont
  {Van~Vleck}},\ }\bibfield  {title} {\bibinfo {title} {{The Jahn-Teller Effect
  and Crystalline Stark Splitting for Clusters of the Form XY$_6$}},\ }\href
  {https://doi.org/10.1063/1.1750327} {\bibfield  {journal} {\bibinfo
  {journal} {J. Chem. Phys.}\ }\textbf {\bibinfo {volume} {7}},\ \bibinfo
  {pages} {72} (\bibinfo {year} {1939})}\BibitemShut {NoStop}%
\bibitem [{\citenamefont {Kanamori}(1960)}]{Kanamori1960CrystalCompounds}%
  \BibitemOpen
  \bibfield  {author} {\bibinfo {author} {\bibfnamefont {J.}~\bibnamefont
  {Kanamori}},\ }\bibfield  {title} {\bibinfo {title} {{Crystal distortion in
  magnetic compounds}},\ }\href {https://doi.org/10.1063/1.1984590} {\bibfield
  {journal} {\bibinfo  {journal} {J. Appl. Phys.}\ ,\ \bibinfo {pages} {14S}}
  (\bibinfo {year} {1960})}\BibitemShut {NoStop}%
\bibitem [{\citenamefont {Goodenough}(1998)}]{Goodenough1998JAHN-TELLERSOLIDS}%
  \BibitemOpen
  \bibfield  {author} {\bibinfo {author} {\bibfnamefont {J.~B.}\ \bibnamefont
  {Goodenough}},\ }\bibfield  {title} {\bibinfo {title} {{Jahn-Teller phenomena
  in solids}},\ }\href {https://doi.org/10.1146/annurev.matsci.28.1.1}
  {\bibfield  {journal} {\bibinfo  {journal} {Annu. Rev. Mater. Sci}\ }\textbf
  {\bibinfo {volume} {28}},\ \bibinfo {pages} {1} (\bibinfo {year}
  {1998})}\BibitemShut {NoStop}%
\bibitem [{\citenamefont {Christy}(1995)}]{Christy1995IsosymmetricExamples}%
  \BibitemOpen
  \bibfield  {author} {\bibinfo {author} {\bibfnamefont {A.~G.}\ \bibnamefont
  {Christy}},\ }\bibfield  {title} {\bibinfo {title} {{Isosymmetric structural
  phase transitions: phenomenology and examples}},\ }\href
  {https://doi.org/10.1107/S0108768195001728} {\bibfield  {journal} {\bibinfo
  {journal} {Acta Crystallogr. B}\ }\textbf {\bibinfo {volume} {51}},\ \bibinfo
  {pages} {753} (\bibinfo {year} {1995})}\BibitemShut {NoStop}%
\bibitem [{\citenamefont {Rodr{\'{i}}guez-Carvajal}\ \emph
  {et~al.}(1998)\citenamefont {Rodr{\'{i}}guez-Carvajal}, \citenamefont
  {Hennion}, \citenamefont {Moussa}, \citenamefont {Moudden}, \citenamefont
  {Pinsard},\ and\ \citenamefont
  {Revcolevschi}}]{Rodriguez-Carvajal1998Neutron-diffractionLaMnO3}%
  \BibitemOpen
  \bibfield  {author} {\bibinfo {author} {\bibfnamefont {J.}~\bibnamefont
  {Rodr{\'{i}}guez-Carvajal}}, \bibinfo {author} {\bibfnamefont
  {M.}~\bibnamefont {Hennion}}, \bibinfo {author} {\bibfnamefont
  {F.}~\bibnamefont {Moussa}}, \bibinfo {author} {\bibfnamefont {A.~H.}\
  \bibnamefont {Moudden}}, \bibinfo {author} {\bibfnamefont {L.}~\bibnamefont
  {Pinsard}},\ and\ \bibinfo {author} {\bibfnamefont {A.}~\bibnamefont
  {Revcolevschi}},\ }\bibfield  {title} {\bibinfo {title} {{Neutron-diffraction
  study of the Jahn-Teller transition in stoichiometric LaMnO$_3$}},\ }\href
  {https://doi.org/10.1103/PhysRevB.57.R3189} {\bibfield  {journal} {\bibinfo
  {journal} {Phys. Rev. B}\ }\textbf {\bibinfo {volume} {57}},\ \bibinfo
  {pages} {R3189} (\bibinfo {year} {1998})}\BibitemShut {NoStop}%
\bibitem [{\citenamefont {Thygesen}\ \emph {et~al.}(2017)\citenamefont
  {Thygesen}, \citenamefont {Young}, \citenamefont {Beake}, \citenamefont
  {Romero}, \citenamefont {Connor}, \citenamefont {Proffen}, \citenamefont
  {Phillips}, \citenamefont {Tucker}, \citenamefont {Hayward}, \citenamefont
  {Keen},\ and\ \citenamefont {Goodwin}}]{Thygesen2017LocalLaMnO3}%
  \BibitemOpen
  \bibfield  {author} {\bibinfo {author} {\bibfnamefont {P.~M.~M.}\
  \bibnamefont {Thygesen}}, \bibinfo {author} {\bibfnamefont {C.~A.}\
  \bibnamefont {Young}}, \bibinfo {author} {\bibfnamefont {E.~O.~R.}\
  \bibnamefont {Beake}}, \bibinfo {author} {\bibfnamefont {F.~D.}\ \bibnamefont
  {Romero}}, \bibinfo {author} {\bibfnamefont {L.~D.}\ \bibnamefont {Connor}},
  \bibinfo {author} {\bibfnamefont {T.~E.}\ \bibnamefont {Proffen}}, \bibinfo
  {author} {\bibfnamefont {A.~E.}\ \bibnamefont {Phillips}}, \bibinfo {author}
  {\bibfnamefont {M.~G.}\ \bibnamefont {Tucker}}, \bibinfo {author}
  {\bibfnamefont {M.~A.}\ \bibnamefont {Hayward}}, \bibinfo {author}
  {\bibfnamefont {D.~A.}\ \bibnamefont {Keen}},\ and\ \bibinfo {author}
  {\bibfnamefont {A.~L.}\ \bibnamefont {Goodwin}},\ }\bibfield  {title}
  {\bibinfo {title} {{Local structure study of the orbital order/disorder
  transition in LaMnO$_3$}},\ }\href
  {https://doi.org/10.1103/PhysRevB.95.174107} {\bibfield  {journal} {\bibinfo
  {journal} {Phys. Rev. B}\ }\textbf {\bibinfo {volume} {95}},\ \bibinfo
  {pages} {174107} (\bibinfo {year} {2017})}\BibitemShut {NoStop}%
\bibitem [{\citenamefont {Goodwin}(2017)}]{Goodwin2017OrbitalTemperatures}%
  \BibitemOpen
  \bibfield  {author} {\bibinfo {author} {\bibfnamefont {A.~L.}\ \bibnamefont
  {Goodwin}},\ }\href {https://goodwingroupox.uk/etc/project-two-xt38h}
  {\bibinfo {title} {{Orbital (dis)order: A tale of two temperatures}}}
  (\bibinfo {year} {2017})\BibitemShut {NoStop}%
\bibitem [{\citenamefont {Herlihy}\ \emph {et~al.}(2025)\citenamefont
  {Herlihy}, \citenamefont {Chen}, \citenamefont {Ritter}, \citenamefont
  {Chuang},\ and\ \citenamefont
  {Senn}}]{Herlihy2025InterplayRuddlesdenPoppers}%
  \BibitemOpen
  \bibfield  {author} {\bibinfo {author} {\bibfnamefont {A.}~\bibnamefont
  {Herlihy}}, \bibinfo {author} {\bibfnamefont {W.-T.}\ \bibnamefont {Chen}},
  \bibinfo {author} {\bibfnamefont {C.}~\bibnamefont {Ritter}}, \bibinfo
  {author} {\bibfnamefont {Y.-C.}\ \bibnamefont {Chuang}},\ and\ \bibinfo
  {author} {\bibfnamefont {M.~S.}\ \bibnamefont {Senn}},\ }\bibfield  {title}
  {\bibinfo {title} {{Interplay between Jahn-Teller Distortions and Structural
  Phase Transitions in Ruddlesden-Poppers}},\ }\href
  {https://doi.org/10.1021/jacs.5c00459} {\bibfield  {journal} {\bibinfo
  {journal} {J. Am. Chem. Soc.}\ }\textbf {\bibinfo {volume} {147}},\ \bibinfo
  {pages} {7209} (\bibinfo {year} {2025})}\BibitemShut {NoStop}%
\bibitem [{\citenamefont {Shanbhag}\ \emph {et~al.}(2023)\citenamefont
  {Shanbhag}, \citenamefont {Fauth},\ and\ \citenamefont
  {Sundaresan}}]{Shanbhag2023ObservationLaVO3}%
  \BibitemOpen
  \bibfield  {author} {\bibinfo {author} {\bibfnamefont {P.~N.}\ \bibnamefont
  {Shanbhag}}, \bibinfo {author} {\bibfnamefont {F.}~\bibnamefont {Fauth}},\
  and\ \bibinfo {author} {\bibfnamefont {A.}~\bibnamefont {Sundaresan}},\
  }\bibfield  {title} {\bibinfo {title} {{Observation of C-type orbital ordered
  phase and orbital flipping in LaVO$_3$}},\ }\href
  {https://doi.org/10.1103/PhysRevB.108.134115} {\bibfield  {journal} {\bibinfo
   {journal} {Phys. Rev. B}\ }\textbf {\bibinfo {volume} {108}},\ \bibinfo
  {pages} {134115} (\bibinfo {year} {2023})}\BibitemShut {NoStop}%
\bibitem [{\citenamefont {Blake}\ \emph {et~al.}(2002)\citenamefont {Blake},
  \citenamefont {Palstra}, \citenamefont {Ren}, \citenamefont {Nugroho},\ and\
  \citenamefont {Menovsky}}]{Blake2002NeutronYVO3}%
  \BibitemOpen
  \bibfield  {author} {\bibinfo {author} {\bibfnamefont {G.~R.}\ \bibnamefont
  {Blake}}, \bibinfo {author} {\bibfnamefont {T.~T.}\ \bibnamefont {Palstra}},
  \bibinfo {author} {\bibfnamefont {Y.}~\bibnamefont {Ren}}, \bibinfo {author}
  {\bibfnamefont {A.~A.}\ \bibnamefont {Nugroho}},\ and\ \bibinfo {author}
  {\bibfnamefont {A.~A.}\ \bibnamefont {Menovsky}},\ }\bibfield  {title}
  {\bibinfo {title} {{Neutron diffraction, x-ray diffraction, and specific heat
  studies of orbital ordering in YVO$_3$}},\ }\href
  {https://doi.org/10.1103/PhysRevB.65.174112} {\bibfield  {journal} {\bibinfo
  {journal} {Phy. Rev. B}\ }\textbf {\bibinfo {volume} {65}},\ \bibinfo {pages}
  {1741121} (\bibinfo {year} {2002})}\BibitemShut {NoStop}%
\end{thebibliography}%

\clearpage
\newpage

\onecolumngrid

\section*{SUPPLEMENTAL MATERIAL}
\setcounter{page}{1}
\setcounter{figure}{0}
\setcounter{table}{0}
\setcounter{section}{0}
\setcounter{equation}{0}
\renewcommand{\thepage}{S\arabic{page}}
\renewcommand{\thesection}{S\arabic{section}}
\renewcommand{\thetable}{S\arabic{table}}
\renewcommand{\thefigure}{S\arabic{figure}}
\newcounter{SIfig}
\renewcommand{\theSIfig}{S\arabic{SIfig}}

\section{Structural Analysis}

\begin{figure}[ht!]
\includegraphics[scale=0.7]{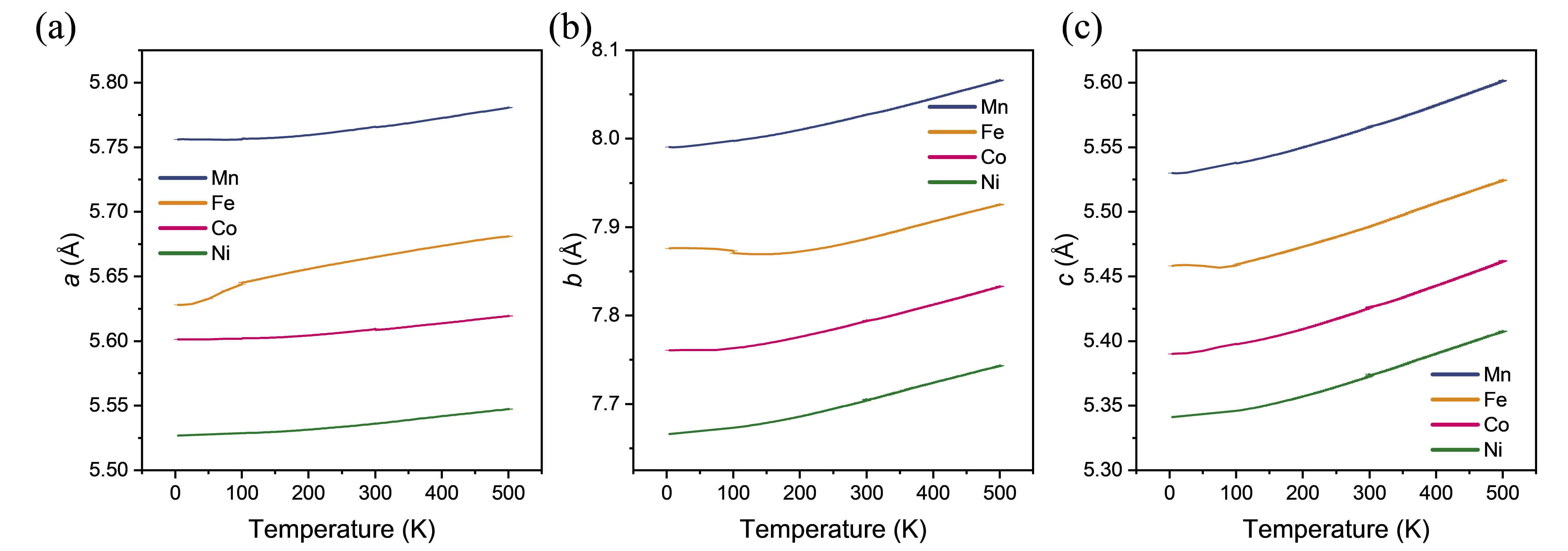}
\caption{\label{fig:S1} Variations in lattice parameters vs temperature.}
\end{figure}

\begin{figure}[ht!]
\includegraphics[scale=0.6]{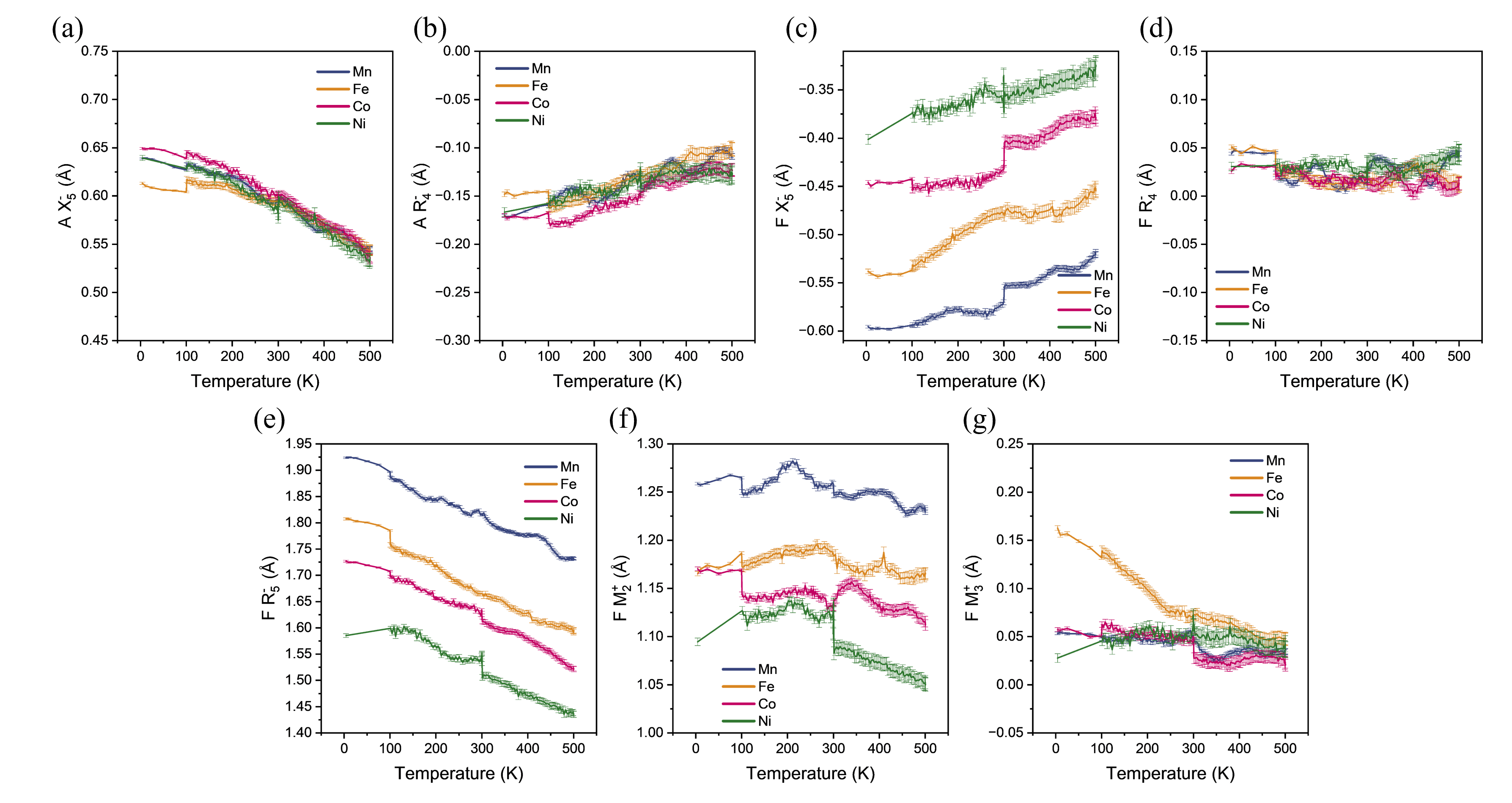}
\caption{\label{fig:S2} Variations in mode amplitude vs temperature for the \textbf{a}. A site X$_5^-$, \textbf{b}. A site R$_4^-$, \textbf{c}. F X$_5^-$, \textbf{d}. F R$_4^-$, \textbf{e}. F R$_5^-$, \textbf{f}. F M$_3^+$ and \textbf{g}. F M$_3^+$ irreps.}
\end{figure}

\begin{table}[ht!]
    \caption{Structural details for NaFeF$_3$ at 300 K obtained from Rietveld refinement against X-ray diffraction data at beamline I11, Diamond Light Source. Atomic positions were refined through symmetry modes generated using ISODISTORT. }
    \begin{tabular}{c c c c c c c}
          \hline \hline
         Atom & Site & \textit{x} & \textit{y} & \textit{z} & Occupancy & \textit{B}$_{iso}$ (Å$^2$) \\ \hline
         \multicolumn{7}{l}{Space Group:\textit{Pnma}, \textit{R}$_{wp}$ = 5.5743 \%, GOF = 4.173 \%  } \\
         \multicolumn{7}{l}{\textit{a} = 5.66304(5) (Å), \textit{b} = 7.8872(7) (Å), \textit{c} = 5.4912(5) (Å), $\alpha=\beta=\gamma$ = 90$^\circ$} \\ \hline
         
        Na(1) & 4\textit{c} & -0.0522(4) & 0.25 & -0.0111(6) & 1 & 1.93(6) \\
        Fe(1) & 4\textit{a} & 0 & 0 & 0 & 1 & 0.77(2) \\
        F(1) & 8\textit{c} & 0.1994(4) & 0.0550(3) & 0.8074(4) & 1 & 1.36(7) \\
        F(2) & 4\textit{c} & 0.0452(6) & 0.25 & 0.3938(6) & 1 & 0.73(9) \\ \hline \hline
    \end{tabular}
    \label{S:table1}
\end{table}

\begin{table}[ht!]
    \caption{Structural details for NaFeF$_3$ at 100 K obtained from Rietveld refinement against X-ray diffraction data at beamline I11, Diamond Light Source. Atomic positions were refined through symmetry modes generated using ISODISTORT. }
    \begin{tabular}{c c c c c c c}
          \hline \hline
         Atom & Site & \textit{x} & \textit{y} & \textit{z} & Occupancy & \textit{B}$_{iso}$ (Å$^2$) \\ \hline
         \multicolumn{7}{l}{Space Group:\textit{Pnma}, \textit{R}$_{wp}$ = 5.3929 \%, GOF = 4.485 \%  } \\
         \multicolumn{7}{l}{\textit{a} = 5.6453(8) (Å), \textit{b} = 7.871(1) (Å), \textit{c} = 5.4592(8) (Å), $\alpha=\beta=\gamma$ = 90$^\circ$} \\ \hline
         
        Na(1) & 4\textit{c} & -0.0561(3) & 0.25 & -0.0147(5) & 1 & 0.86(4) \\
        Fe(1) & 4\textit{a} & 0 & 0 & 0 & 1 & 0.25(2) \\
        F(1) & 8\textit{c} & 0.2032(3) & 0.0571(2) & 0.8094(3) & 1 & 0.25(4) \\
        F(2) & 4\textit{c} & 0.0483(5) & 0.25 & 0.3889(4) & 1 & 0.37(6) \\ \hline \hline
    \end{tabular}
    \label{S:table2}
\end{table}

\begin{table}[ht!]
    \caption{Structural details for NaFeF$_3$ at 4 K obtained from Rietveld refinement against X-ray diffraction data at beamline ID22, ESRF. Atomic positions were refined through symmetry modes generated using ISODISTORT.}
    \begin{tabular}{c c c c c c c}
          \hline \hline
         Atom & Site & \textit{x} & \textit{y} & \textit{z} & Occupancy & \textit{B}$_{iso}$ (Å$^2$) \\ \hline
         \multicolumn{7}{l}{Space Group:\textit{Pnma}, \textit{R}$_{wp}$ = 4.9679 \%, GOF = 0.974 \%  } \\
         \multicolumn{7}{l}{\textit{a} = 5.62787(2) (Å), \textit{b} = 7.87607(2) (Å), \textit{c} = 5.45820(2) (Å), $\alpha=\beta=\gamma$ = 90$^\circ$} \\ \hline
         
        Na(1) & 4\textit{c} & -0.0555(1) & 0.25 & -0.0133(2) & 1 & 0.59(1) \\
        Fe(1) & 4\textit{a} & 0 & 0 & 0 & 1 & 0.113(4) \\
        F(1) & 8\textit{c} & 0.2042(2) & 0.0598(1) & 0.8104(2) & 1 & 0.32(1) \\
        F(2) & 4\textit{c} & 0.0489(2) & 0.25 & 0.3868(2) & 1 & 0.34(2) \\ \hline \hline
    \end{tabular}
    \label{S:table3}
\end{table}

\newpage

\begin{figure}[ht!]
\includegraphics[scale=1.5]{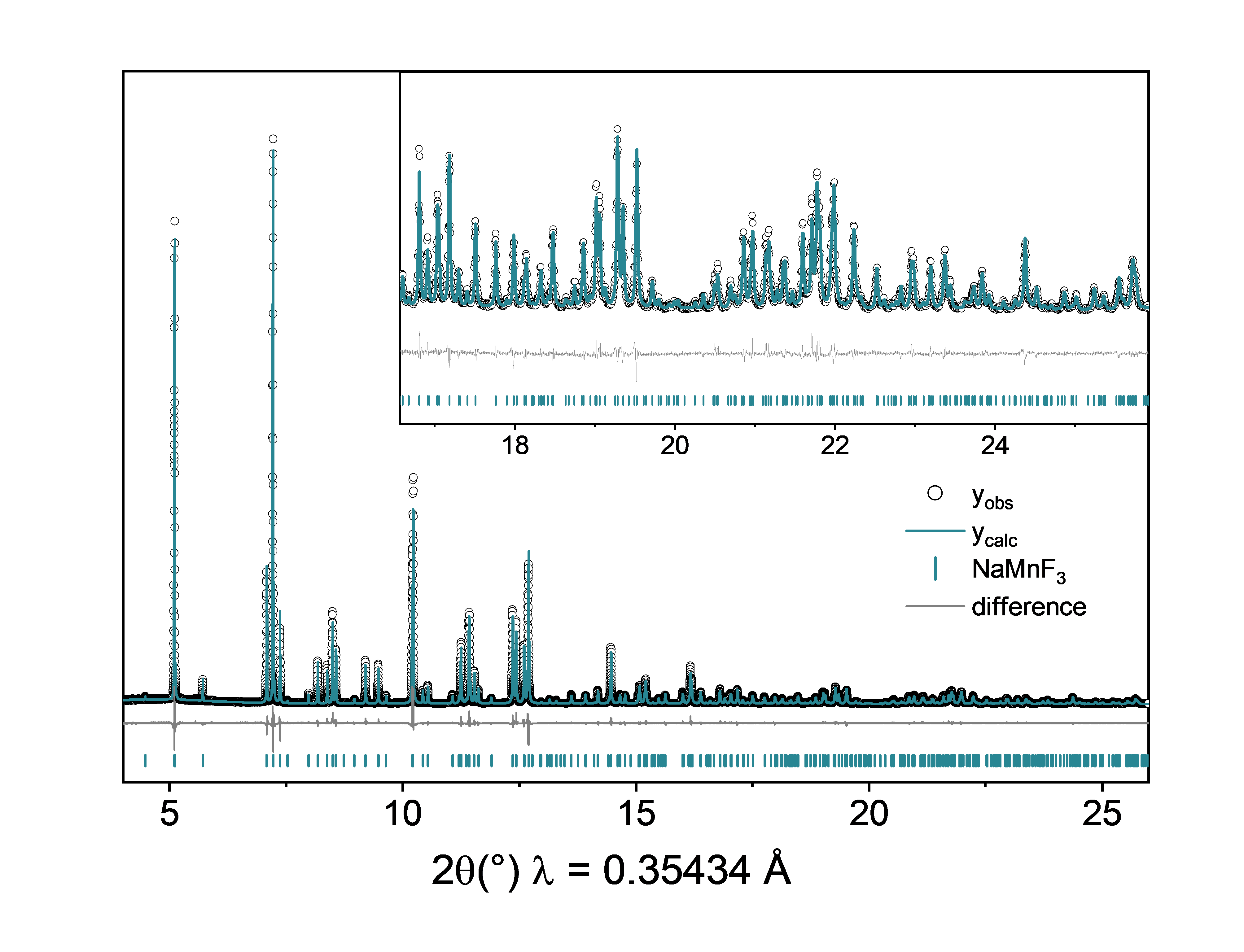}
\caption{\label{fig:S-riet1} Rietveld refinement of NaMnF$_3$ in \textit{Pnma} on data collected at 10 K on ID22.}
\end{figure}

\begin{figure}[ht!]
\includegraphics[scale=1.5]{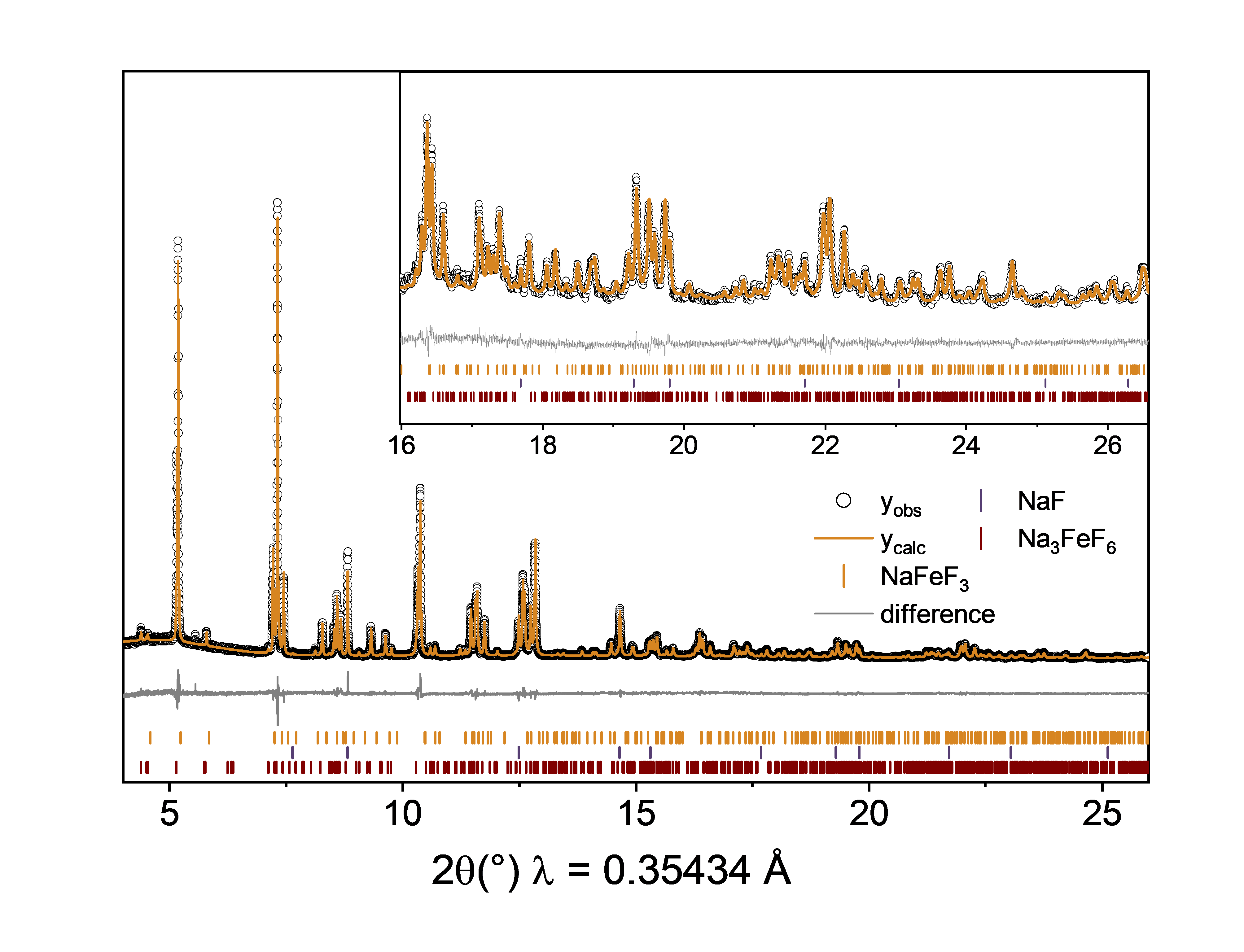}
\caption{\label{fig:S-riet2} Rietveld refinement of NaFeF$_3$ in \textit{Pnma} on data collected at 10 K on ID22.}
\end{figure}

\begin{figure}[ht!]
\includegraphics[scale=1.5]{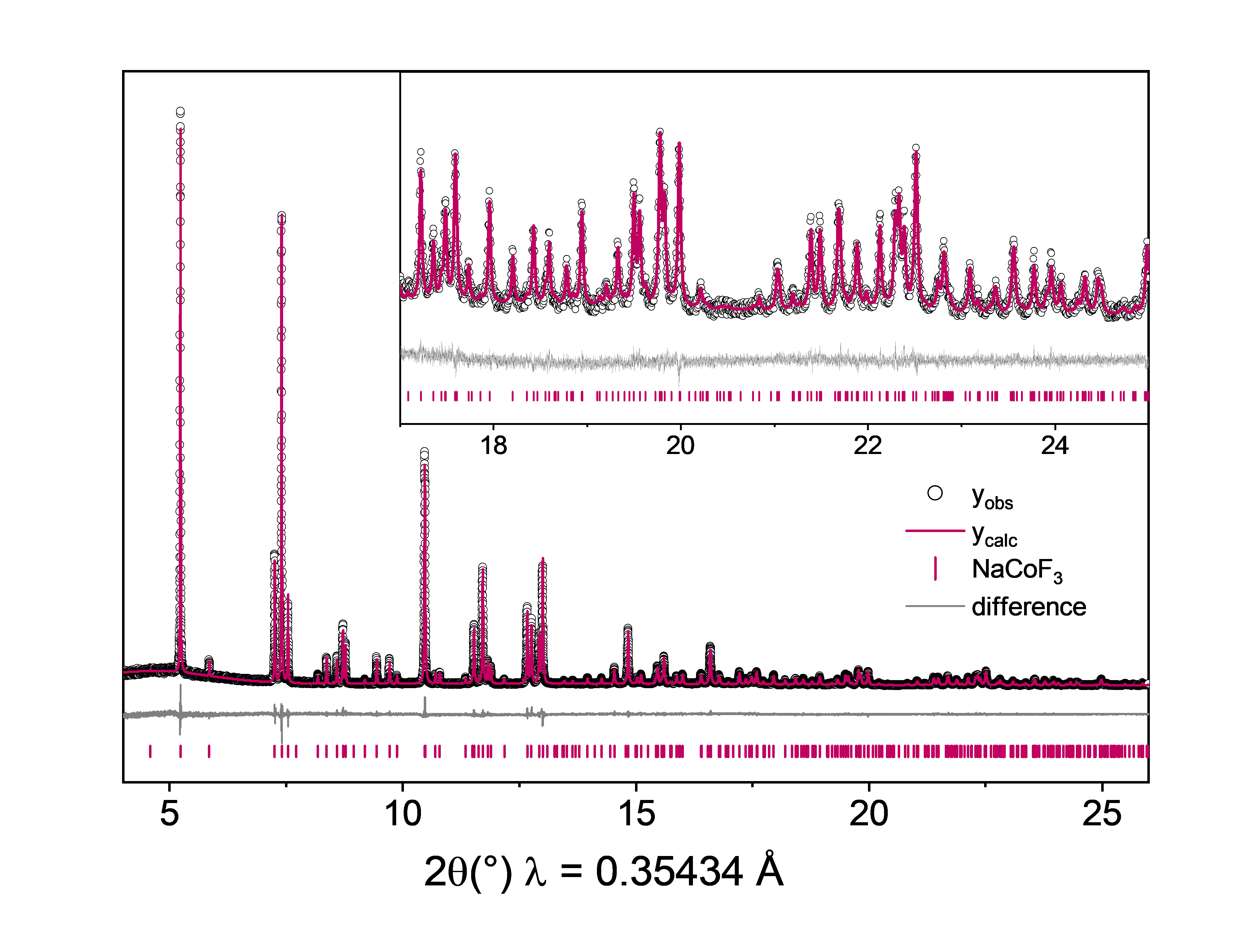}
\caption{\label{fig:S-riet3} Rietveld refinement of NaCoF$_3$ in \textit{Pnma} on data collected at 10 K on ID22.}
\end{figure}

\begin{figure}[ht!]
\includegraphics[scale=1.5]{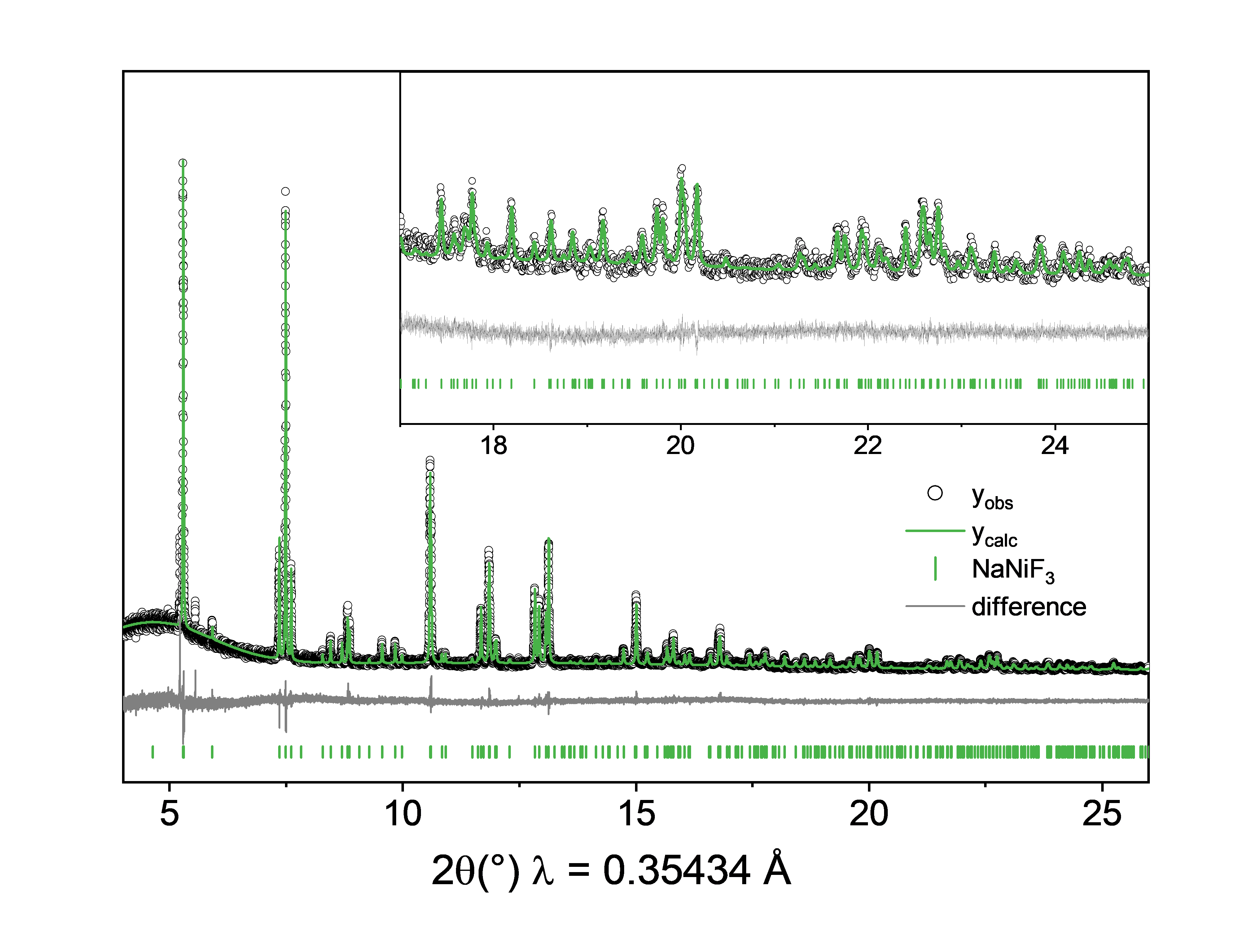}
\caption{\label{fig:S-riet4} Rietveld refinement of NaNiF$_3$ in \textit{Pnma} on data collected at 10 K on ID22.}
\end{figure}

\clearpage
\newpage

\section{Details on Calculation of Van Vleck distortion Modes}
Van Vleck distortion modes were calculated using Python package \textit{VanVleckCalculator} to allow the ease of calculation of the bond angle distortions; \textit{Q$_{4,5,6}$}. Bond length distortions (\textit{Q$_{2,3}$}) where calculated both by hand using bond lengths, and using \textit{VanVleckCalculator} and gave results that were in agreement with each other. 

For this an \textit{xyz} coordinate system was chosen with respect to the octahedron, where \textit{z} is roughly along \textit{b} of the \textit{Pnma} unit cell, and \textit{x,y} are in \textit{ac}. These axes were chosen for ease of comparison to the well studied LaMnO$_3$ \textit{Pnma} perovskites. This formulation means that in the case of a purely tetragonal distortion of the unit cell and octahedra (e.g. $\Gamma_3^+$) the tetragonal compression/elongation will give a pure \textit{Q$_{3z}$} motion.   

For the calculation of \textit{Q$_{2,3}$} from bond lengths, the following equations were used: 

\begin{equation}
    Q_2 = \frac{2}{\sqrt{2}}(x-y)
    \label{placeholder}
\end{equation}
\begin{equation}
    Q_3 = \frac{2}{\sqrt{6}}(2z - x - y)
    \label{placeholder}
\end{equation}
where \textit{x,y,z} denote the bond lengths along the specified octahedron axes. 

\textit{Q$_{4,5,6}$} were calculated using a Cartesian coordinates method using \textit{VanVleckCalculator}. A description of the method is provided in \cite{Nagle-Cocco2024VanVanVleckCalculator}.  

\begin{figure}[ht!]
\includegraphics[scale=0.6]{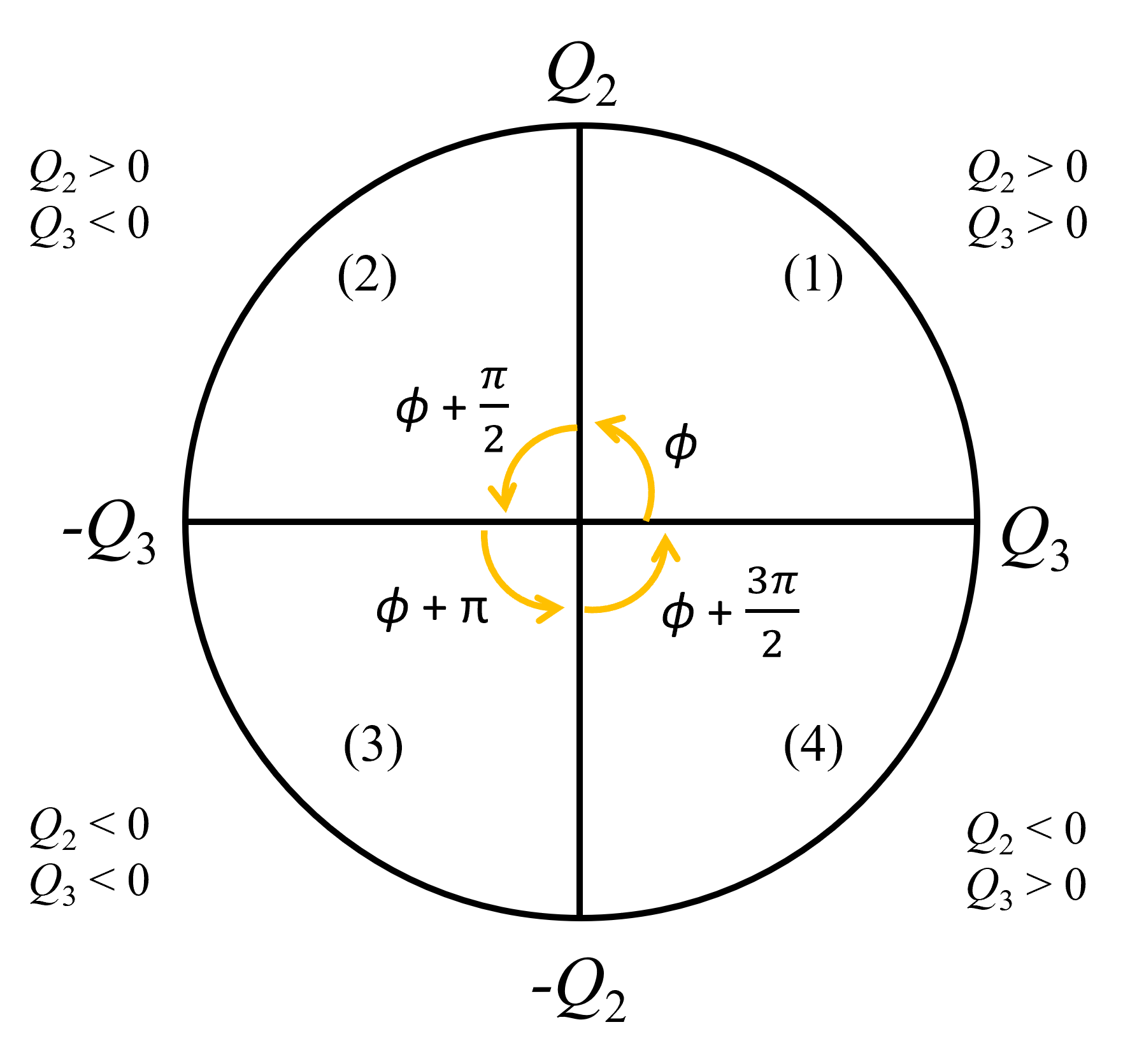}
\caption{\label{fig:S3} Polar plot of \textit{Q}$_2$-\textit{Q}$_3$ distortion space highlighting the four quadrants.}
\end{figure}

\begin{figure}[ht!]
\includegraphics[scale=0.9]{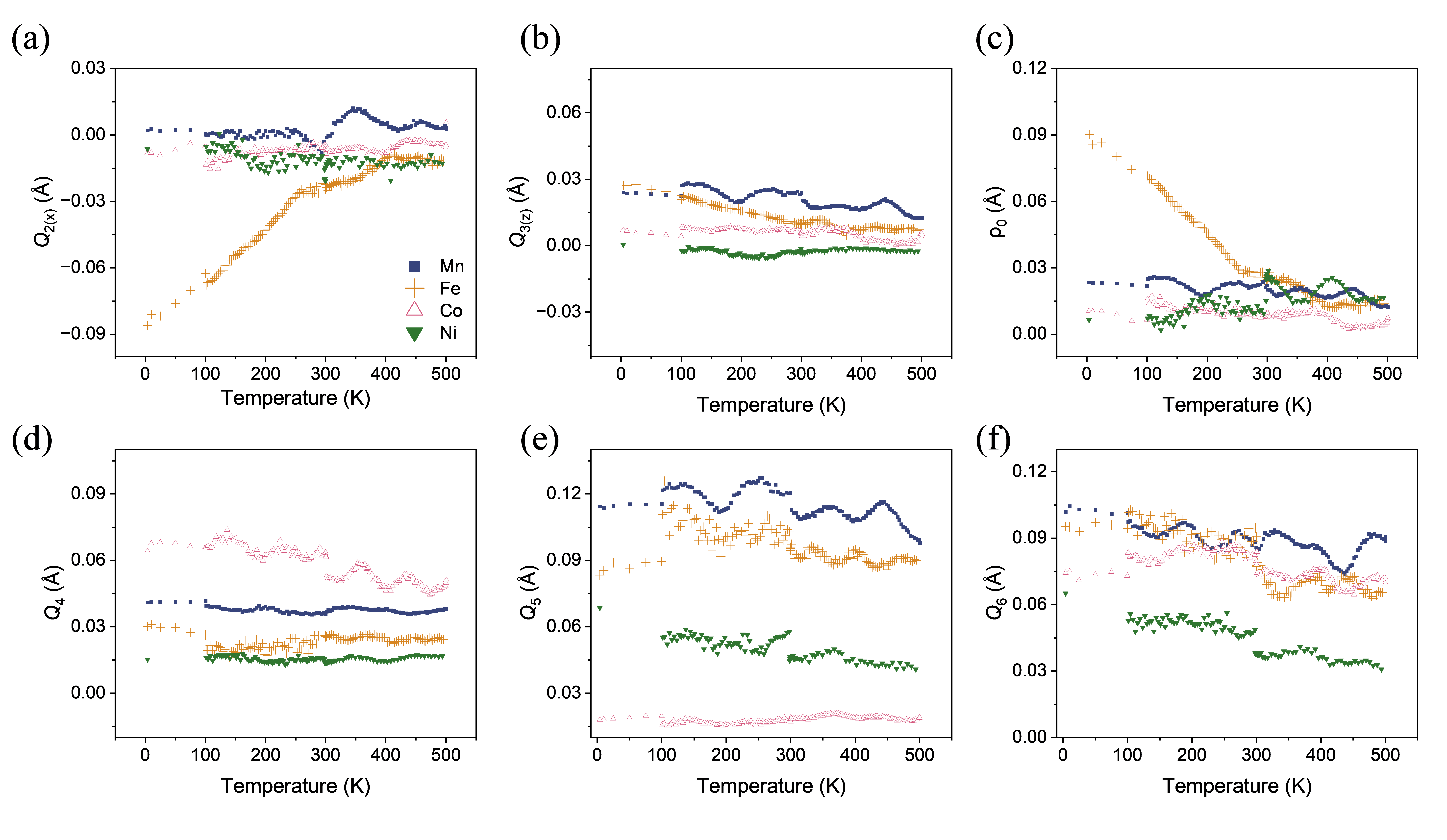}
\caption{\label{fig:S4} Van Vleck distortion modes and $\rho_0$ verses temperature. \textit{Q$_{4,5,6}$} remain relatively constant with decreasing temperature indicating that the angular distortion likely results from structural origins such as octahedral tilting rather than the onset of any electronic origins.}
\end{figure}

\begin{figure}[ht!]
\includegraphics[scale=0.9]{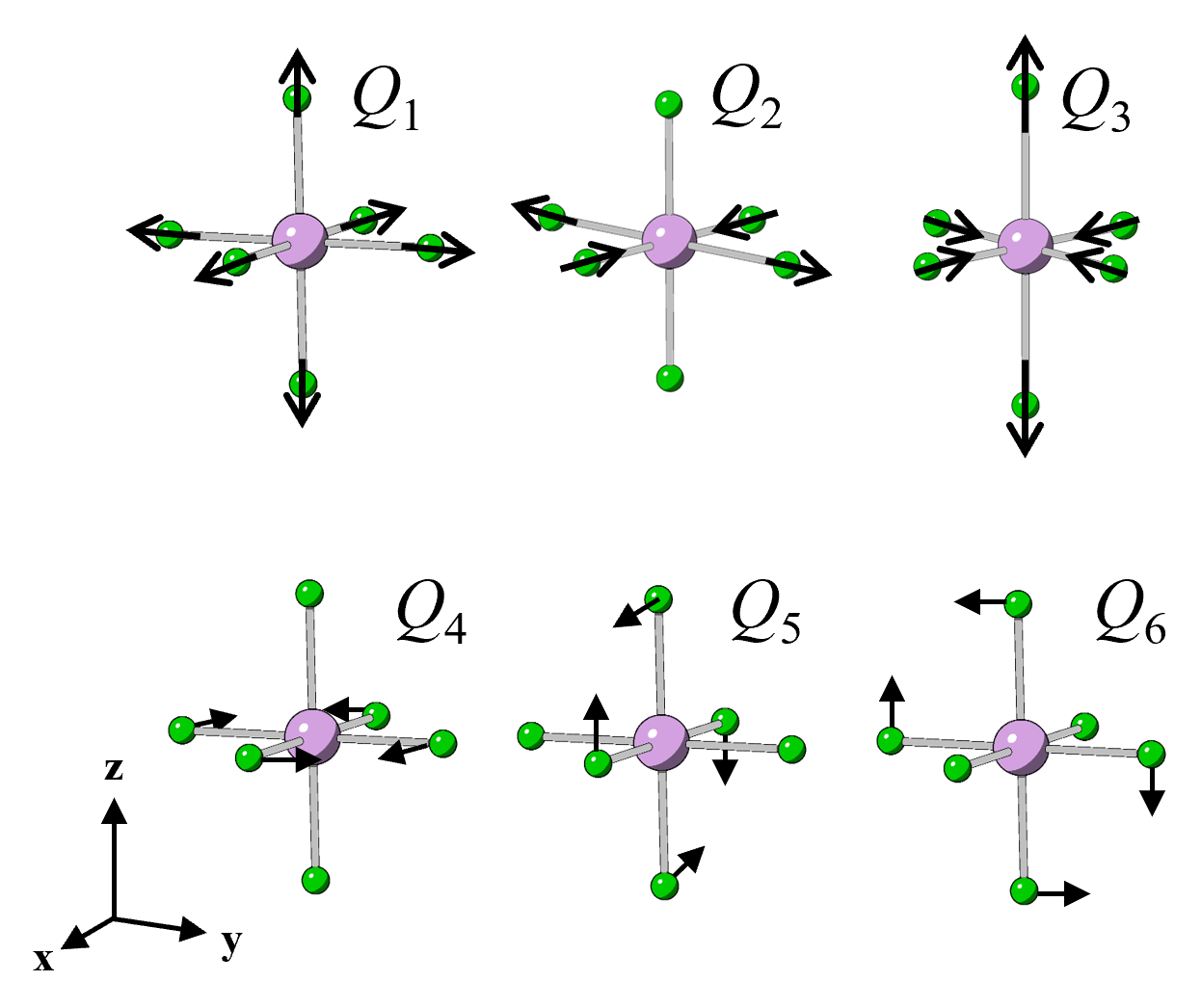}
\caption{\label{fig:S5} Bond length distortions (\textit{Q}$_{1,2,3}$) and bond angle distortions (\textit{Q}$_{4,5,6}$) associated with the Van Vleck distortion modes.}
\end{figure}

\begin{figure}[ht!]
\includegraphics[scale=0.9]{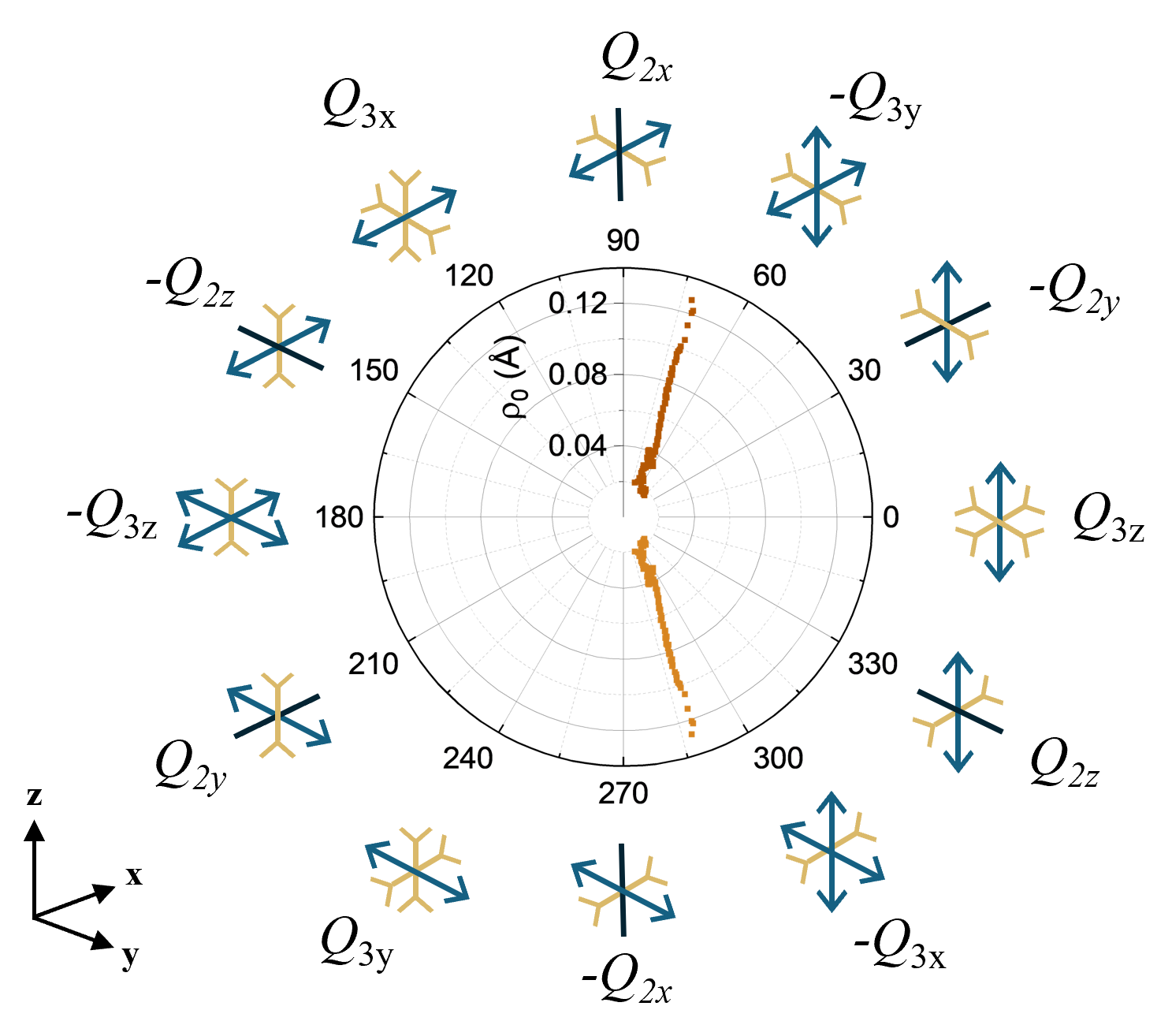}
\caption{\label{fig:s6} Polar plot of \textit{Q}$_2$-\textit{Q}$_3$ distortion space with $\rho_0$ and $\phi$ values calculated for the two neighbouring FeF$_6$ octahedra related by a rotation of 90$^{\circ}$. The negative \textit{Q}$_3$ compressions along the \textit{x} and \textit{y} axes are indicative of doubly occupied \textit{d$_{yz}$} or \textit{d$_{xz}$} orbitals with respect to the neighbouring octahedra.   }
\end{figure}

To calculate $\phi$ correctly in \textit{Q}$_2$-\textit{Q}$_3$ distortion space, additional requirements needed to be made to ensure that the correct quadrant was chosen (Fig.\ \ref{fig:S3}) with respect to the arctan of the ratio of \textit{Q}$_2$ to \textit{Q}$_3$. This was done by looking at the values of \textit{Q}$_2$ and \textit{Q}$_3$ to identify which quadrant $\phi$ would result in.

The product of \textit{Q}$_2$\textit{Q}$_3$ was also used to identify the quadrant depending on whether the value was greater or less than 0. 
\begin{eqnarray}
   & Q_2Q_3 < 0 ;\ use\ \phi = atan(Q_3/Q_2)\\
   & Q_2Q_3 > 0;\ use\ \phi = atan(Q_2/Q_3) 
\end{eqnarray}
A positive value indicates that \textit{Q}$_2$ and \textit{Q}$_3$ are either both positive (quadrant 1) or both negative (quadrant 3) and equation 4 is used to calculate the angle away from \textit{Q}$_3$, whereas if the value is negative then \textit{Q}$_2$ and \textit{Q}$_3$ have opposing signs and equation 3 is used to calculate the angle away from \textit{Q}$_2$. To account for the limits of an arctangent (-$\pi$/2 to $\pi$/2) a factor of $\pi$ was added to the calculated value as detailed in the figure.

\newpage
\section{Van Vleck Analysis and Polar Plots}

\clearpage
\newpage
\section{INVARIANTS ANALYSIS}

\begin{figure}[ht!]
\includegraphics[scale=0.8]{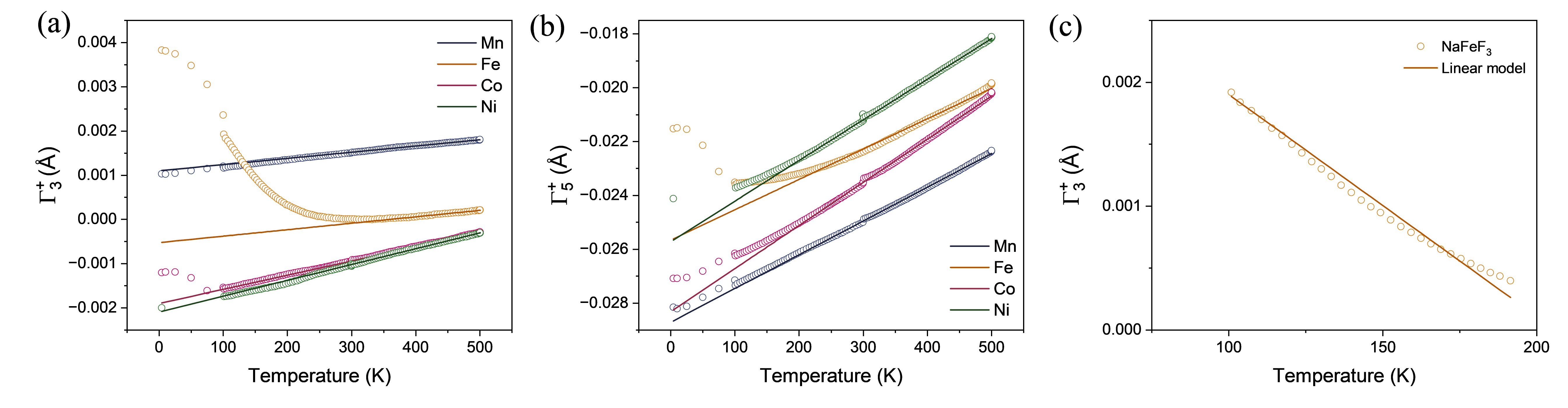}
\caption{\label{fig:s7} Linear models determined from INVARIANTS analysis of the (\textbf{a}.) $\Gamma_3^+$ and (\textbf{b}.) $\Gamma_5^+$  strain modes. Figures a. and b. are modelled using contributions from only M$_2^+$ and R$_5^-$ modes. \textbf{c}. Shows the low temperature region of $\Gamma_3^+$ with the M$_3^+$ mode included in the fitting. }
\end{figure}

INVARIANTS analysis was used to investigate the coupling schemes present in the \textit{Pnma} perovskites and identify the main terms that couple the internal degrees of freedom to symmetry breaking strain in the structure. Through use of INVARIANTS\cite{Hatch2003INVARIANTS:Group}\cite{StokesINVARIANTSIso.byu.edu.}, we can ascertain how the M$_2^+$, R$_5^-$ and M$_3^+$ modes couple together to give the $\Gamma_5^+$ and $\Gamma_3^+$ strain responses. 

\begin{equation}
\Gamma_3^+ = \alpha(M_2^+)^2 + \beta(M_3^+)^2 + \gamma(R_5^-)^2 
\end{equation}
and 
\begin{equation}
\Gamma_5^+ = \alpha(M_2^+ M_3^+) + \beta(R_5^-)^2 
\end{equation}

For M = Mn, Co and Ni at all temperatures, and M = Fe above 400 K, $\Gamma_3^+$ can be modelled with contributions from only the M$_2^+$ and R$_5^-$ modes without the need for the M$_3^+$ mode (which is very small and near constant within these temperature regimes). Table \ref{S:table4} and Table \ref{S:table5} detail all fitting parameters used. To reduce noise incurring from multiplying through small variations in mode amplitude, simultaneous straight line fits of the individual modes were used in place of the values output from refinement. This followed the form \(y = (m_{irrep}\times T) + (c_{irrep}\times\) offset factor) where \textit{T} is the temperature, \(m_{irrep}\) and \(c_{irrep}\) were global parameters and \textit{offset factor} was a refined factor to account for changes in the tilt of the octahedra due to different ionic radii. $\alpha$ and $\gamma$ varied slightly between compositions to account for the difference in ionic radii. 

Fig.\ \ref{fig:s7}(a) shows the resulting fitting to the $\Gamma_3^+$ strain mode. For M = Mn and Ni, the fits without contributions from the M$_3^+$ mode fit well at all temperatures, and NaCoF$_3$ fits well until a small deviation below 100 K at the onset of magnetic order. NaFeF$_3$ however deviates significantly away from this model well above any magnetic order. Inclusion of the M$_3^+$ mode at intermediate temperatures between 100 K and 190 K enables the $\Gamma_3^+$ strain to be modelled in NaFeF$_3$, signifying the coupling of C-type orbital order to the tetragonal strain (Fig.\ \ref{fig:s7}(c)). 

The same analysis can be performed on the $\Gamma_5^+$ strain modes, which can all be modelled by the R$_5^-$ and M$_2^+$ modes alone, and the M$_3^+$ mode is not required to model the shearing strain (see Fig.\ \ref{fig:s7}(b) and Table\ \ref{S:table5}). Any deviations from the model appear to coincide with the onset of magnetic ordering and will likely be resultant of magnetostriction. 

\begin{table}[ht!]
    \caption{Global Parameters used for the simultaneous straight line fitting of M$_2^+$ and R$_5^-$ mode amplitudes versus temperature with \(y = (m_{mode}\times T) + (c_{mode}\times\)offset factor), and individual parameters used for simultaneous fitting of the $\Gamma_3^+$ mode from straight line mode fits at high temperatures. Fe(HT) represents extracted parameters from fitting above 450 K without the inclusion of the M$_3^+$ mode whereas Fe(IT) represents the values extracted from the inclusion of the M$_3^+$ mode with a fit to the 100-180 K temperature region. }
    \begin{tabular}{c|c c c c c}
          \hline \hline
         \multicolumn{6}{c}{$\Gamma_3^+$ = $\alpha$(M$_2^+)^2$ + $\beta$(M$_3^+)^2$ + $\gamma$(R$_5^-)^2$} \\[1.5ex] \hline
        Global Parameters & \multicolumn{5}{c}{} \\ \hline
        m$_{R5-}$ & \multicolumn{5}{|c}{-0.00138(1)} \\ 
        c$_{R5-}$ & \multicolumn{5}{|c}{4.108(3)} \\ 
        m$_{M2+}$ & \multicolumn{5}{|c}{-0.00027(4)} \\ 
        c$_{M2+}$ & \multicolumn{5}{|c}{1.884(5)} \\ \hline
         Individual Parameters        & Mn          & Fe(HT)       & Fe(IT)    & Co  & Ni    \\ \hline
        $\alpha$      & 0.005(11)     & 0.003(7)    & 0.0027(94)   & 0.006(1)  & 0.007(1)   \\ 
        $\beta$       & 0             & 0           & 0.21(16)     & 0         & 0   \\ 
        $\gamma$      & -0.002(5)     & -0.0017(40) & -0.0020(40)  & -0.003(5) & -0.004(6)   \\ 
        offset factor\footnotemark[1] & 0.8947(6)   & 0.7817(9)    & 0.7817(9) & 0.7375(4) & 0.6669(5) \\ \hline \hline
    \end{tabular}
    \footnotetext[1] {offset factor was used to account for variations in tilting due to differing ionic radii.}
    \label{S:table4}
\end{table}

\begin{table}[ht!]
    \caption{Global Parameters used for the simultaneous straight line fitting of M$_2^+$ and R$_5^-$ mode amplitudes versus temperature with \(y = (m_{mode}\times T) + (c_{mode}\times\)offset factor), and individual parameters used for simultaneous fitting of the $\Gamma_5^+$ mode from straight line mode fits at high temperatures. For all compositions the M$_3^+$ mode was not included in the analysis. }
    \begin{tabular}{c|c c c c}
          \hline \hline
         \multicolumn{5}{c}{$\Gamma_5^+$ = $\alpha$(M$_2^+$M$_3^+$) + $\beta$(R$_5^-)^2$ } \\[1.5ex] \hline
        Global Parameters & \multicolumn{4}{c}{} \\ \hline
        m$_{R5-}$ & \multicolumn{4}{|c}{-0.00138(1)} \\ 
        c$_{R5-}$ & \multicolumn{4}{|c}{4.108(3)} \\ 
        m$_{M2+}$ & \multicolumn{4}{|c}{-0.00013(8)} \\ 
        c$_{M2+}$ & \multicolumn{4}{|c}{1.582(2)} \\ \hline
        Individual Parameters & Mn     & Fe & Co  & Ni    \\ \hline
        $\alpha$    & 0.005(32)        & 0.0006(29)      & 0.008(30)   & 0.005(24)   \\ 
        $\beta$     & -0.009(12)       & -0.008(13)     & -0.012(11)   & -0.011(11)   \\ 
        offset factor\footnotemark[1]  & 0.8947(6)   & 0.7817(9)  & 0.7375(4)  & 0.6669(5)  \\ 
        \hline \hline
    \end{tabular}
    \footnotetext[1] {offset factor was used to account for variations in tilting due to differing ionic radii.}
    \label{S:table5}
\end{table}

\clearpage
\newpage
\section{Identification of orbital ordering temperature from a critical temperature fit}

\begin{figure}[ht!]
\includegraphics[scale=0.9]{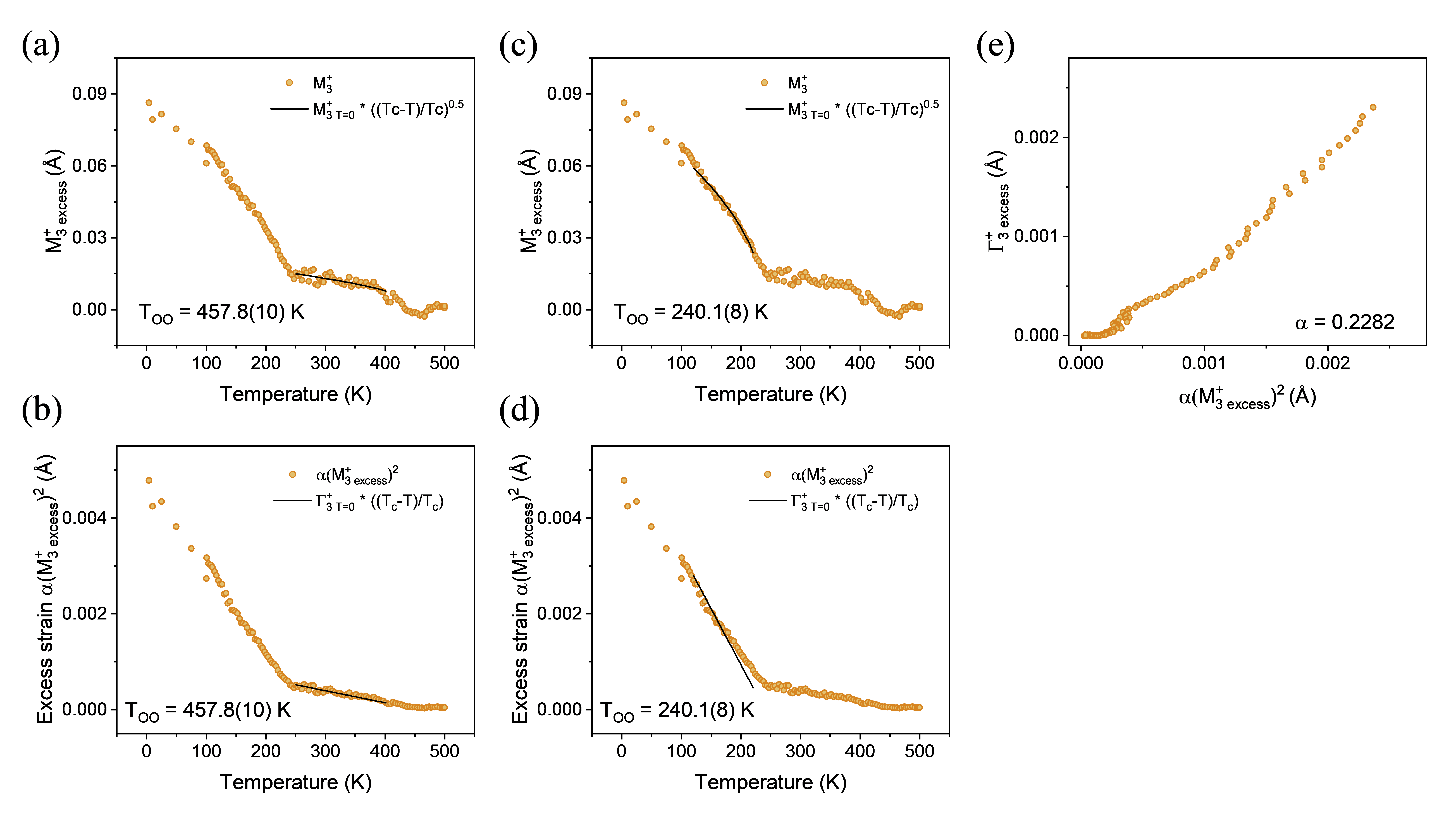}
\caption{\label{fig:s8} Fitting of the \textbf{a}. excess M$_3^+$ and \textbf{b}. excess strain to equations (5) and (6) respectively over the high temperature region to give a T$_c$ = 457.8(10) K. \textbf{c}. and \textbf{d}. show the equivalent model over the lower temperature region to give a T$_c$ = 240.1(8) K. \textbf{e}. The linearly proportional relationship between the excess $\Gamma_3^+$ strain and excess (M$_3^+)^2$ where $\alpha$ is a scaling factor.}
\end{figure}

To identify the temperature at which orbital ordering occurs in NaFeF$_3$ critical temperature fits were performed. In a system where the strain is coupled to an order parameter or specific structural distortion (in this case the M$_3^+$) the excess $\Gamma_3^+$ strain arising from beyond structural origins (\textit{i.e.} electronic origins) should scale proportionally to $\alpha$(M$_3^+)^2$. Since the M$_3^+$ mode is intrinsic to \textit{Pnma} without requiring any JT distortion present, the average magnitude (0.04801 Å) above 450 K is subtracted to provide the excess M$_3^+$ arising from the JT distortion. A similar preparation of the excess $\Gamma_3^+$ strain is done by subtracting the $\Gamma_3^+$ values from a model in which there is no orbital order identified through the INVARIANTS analysis above. We then find that the excess $\Gamma_3^+$ and excess M$_3^+$ scale proportionally (Fig.\ \ref{fig:s8}(e)) highlighting the coupling of the secondary order parameter, M$_3^+$ to the tetragonal strain. 

In a second order phase transition, simultaneously fitting a critical temperature fit to the distortion mode responsible for the transition (the excess M$_3^+$) to equation (5) and the excess strain (excess $ \Gamma_3^+ = (\alpha \ \times$ excess M$_3^+$)) to equation (6) can be used to extract a transition temperature. 

\begin{eqnarray}
   & M_{3~excess}^+(T) = M_{3~excess(T=0)}^+\sqrt{\frac{T_c-T}{T_c}}  \\
   & \Gamma_{3~excess}^+(T) = \Gamma_{3~excess(T=0)}^{+}\frac{T_c-T}{T_c}
\end{eqnarray}

\clearpage
\newpage
\section{Magnetometry and Magnetic structures}
\begin{figure}[ht!]
\includegraphics[scale=0.75]{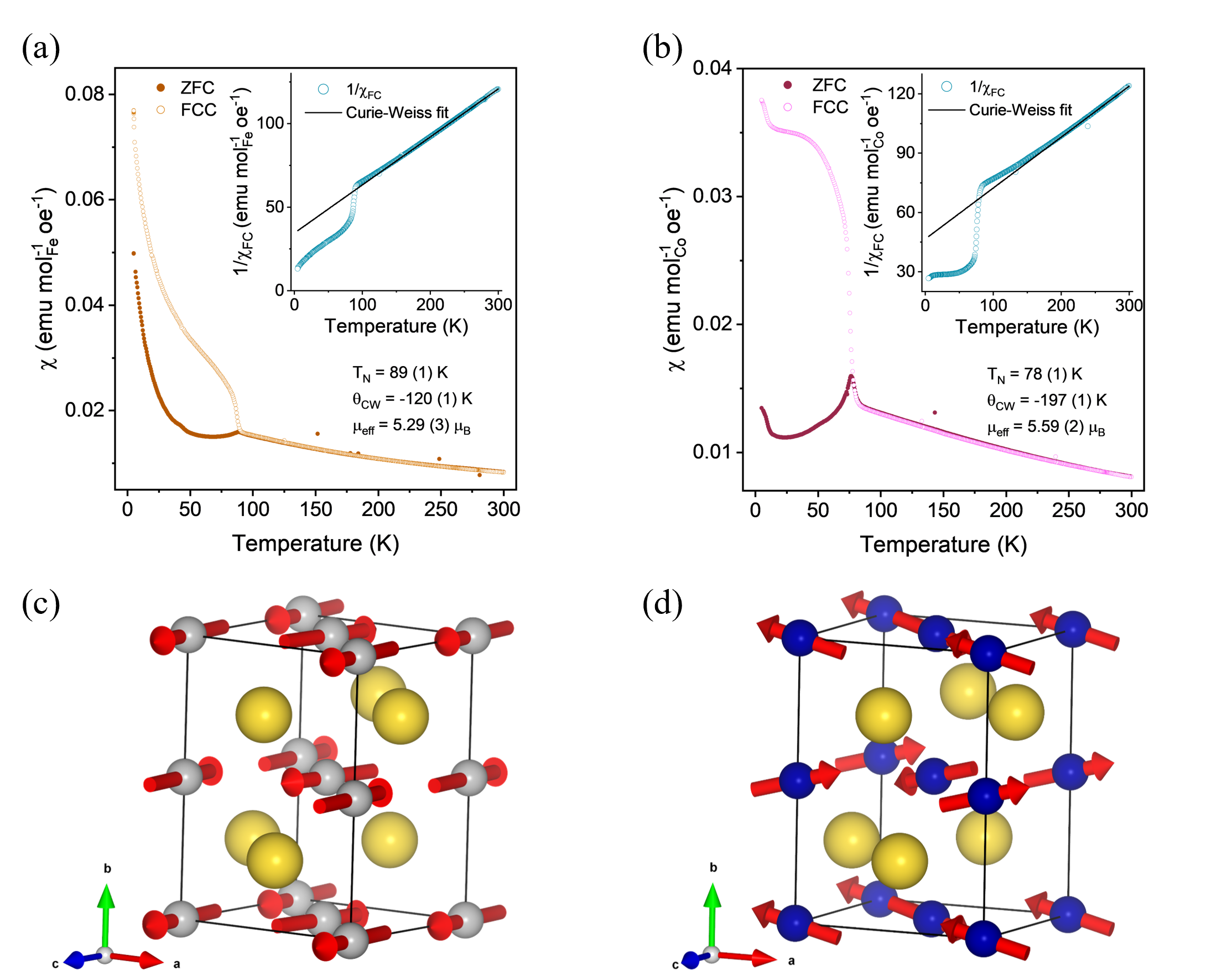}
\caption{\label{fig:s9} M(T) measurements for \textbf{a} NaFeF$_3$ and \textbf{b} NaCoF$_3$. ZFC (warming) data are represented by the darker, filled circles and FC (cooling) data are represented by lighter, hollow circles. Insets show the inverse susceptibility and an extrapolated Curie-Weiss fit to the high temperature paramagnetic region (fitting performed above 250 K). \textbf{c} The magnetic structure for NaFeF$_3$, \textit{Pn}$^\prime$\textit{ma}$^\prime$ (BNS number 62.448), described by the m$\Gamma$$_{4}^{+}$ representation with AFM moments directed along \textit{c}, and weak FM canting along \textit{a}. \textbf{d} The magnetic structure for NaCoF$_3$, \textit{Pnma}.1 (BNS number 62.441) described by the m$\Gamma$$_{1}^{+}$ representation. }
\end{figure}

 Both NaFeF$_3$ and NaCoF$_3$ show bulk AFM transitions (indicated by the negative $\theta_{CW}$ values) accompanied by a weak ferromagnetic (FM) component as evidenced by the splitting of the field cooled (FC) and zero-field cooled (ZFC) measurements (Fig.\ \ref{fig:s9}(a,b)). Curie-Weiss fitting of the paramagnetic region for each composition yielded $\mu_\textit{{eff}}$ values that were indicative of both TM being in their high-spin electron configuration and are in line with previous reports. This is important when comparing NaFeF$_3$ to LnCoO$_3$, which are both \textit{d$^6$}, as the Co$^{3+}$ tends to form the LS, \textit{t$_{2g}^6$} electron configuration which is not JT active. 
The magnetic structure of NaFeF$_3$ has been thoroughly studied and has been determined to have the same magnetic structure as the Mn and Ni analogues, \textit{Pn}$^\prime$\textit{ma}$^\prime$ (BNS number 62.448). This consists of a large G-type AFM along \textit{c}, with a small A-type AFM component along \textit{a}, and weak FM spin canting allowed along \textit{b}. NaCoF$_3$ however, required a different magnetic subgroup, \textit{Pnma}.1 (BNS number 62.441) to allow magnetic components along all three crystallographic axes. The observed wFM cannot be accounted for by the \textit{Pnma}.1 space group, but would be consistent with a lowering of magnetic symmetry to \textit{P}2$_1$/\textit{c}.1. However, since this does not improve the quality of the fit, and we cannot unambiguously determine that the observed macroscopic FM is intrinsic, we choose not to lower the symmetry of our model and the weak AFM canting along \textit{c} was fixed to 0.

\begin{table}[ht!]
    \caption{A comparison of the magnetic structures for NaFeF$_3$ from \cite{Bernal2020CantedMethod} and NaCoF$_3$ from Rietveld refinement on powder neutron diffraction data.}
    \begin{tabular}{c c c c c c c }
          \hline \hline
          & Magnetic SG & BNS number & \textit{M$_x$} ($\mu_B$/FU\footnotemark[1]) & \textit{M$_y$} ($\mu_B$/FU) & \textit{M$_z$} ($\mu_B$/FU) & \textit{M$_{total}$} ($\mu_B$/FU) \\ \hline

        Fe & \textit{Pn$^{\prime}$ma$^{\prime}$} & 62.448 & 0.422(11) & 0\footnotemark[2] & 4.221(4) & 4.242(12) \\
        Co & \textit{Pnma}.1 & 62.441 & -2.77(2) & 0.75(4) & 0\footnotemark[2] & 2.88(7) \\ \hline \hline
    \end{tabular}
    \footnotetext[1] {where FU defines formula unit.}
    \footnotetext[2] {\textit{M$_y$} is fixed at 0 for refinements in \cite{Bernal2020CantedMethod} and for \textit{M$_z$} in this work.}
    \label{S:table7}
\end{table}

\end{document}